\documentclass[]{gensi}
\usepackage{tabularx}
\usepackage[toc,page,header]{appendix}

\usepackage{minitoc}
\usepackage{cleveref} 
\usepackage{microtype}
\usepackage{subcaption}
\usepackage{amsmath} 
\usepackage{amssymb} 
\usepackage{amsfonts}       
\usepackage{pgfplots}
\usepackage{pgfplotstable}
\usepackage{xcolor}
\usepackage{CJKutf8}
\usetikzlibrary{patterns}
\usepackage{multirow}
\usepackage{setspace}
\usepackage{xcolor}
\usepackage{hyperref}
\tcbuselibrary{theorems}
\usepackage{float}
\usepackage{caption}
\usepackage{wrapfig}
\usepackage{xspace}
\usepackage{graphicx}      
\usepackage{subcaption}    

\usepackage{tcolorbox}
\newtcolorbox{DefinitionBox}{
  colback=blue!5,
  colframe=blue!80,
  boxrule=0.5pt,
  arc=2pt,
  left=2pt,
  right=2pt,
  top=2pt,
  bottom=2pt,
}

\newtcolorbox{CorollaryBox}{
  colback=gray!5,
  colframe=gray!80,
  boxrule=0.5pt,
  arc=2pt,
  left=2pt,
  right=2pt,
  top=2pt,
  bottom=2pt,
}
\usepackage{fix-cm}

\usepackage{natbib}
\usepackage{latexsym}

\usepackage{url}
\usepackage{amssymb}
\usepackage[utf8]{inputenc}
\usepackage{microtype}
\usepackage{booktabs}
\usepackage{pifont} 
\usepackage{multirow}
\usepackage{makecell}
\usepackage{paralist}
\usepackage{xspace}
\usepackage{color}
\usepackage{xcolor}
\usepackage{colortbl}
\usepackage{adjustbox}
\usepackage{hyperref} 
\usepackage[edges]{forest}
\usepackage{tikz} 
\usepackage{caption}
\usepackage{amsfonts}
\usepackage{tcolorbox}
\usepackage{algorithm}
\usepackage{algpseudocode}
\usepackage{amsthm}

\usepackage{mathtools}
\usepackage[version=4]{mhchem}
\usepackage{array}  
\usepackage{graphicx}

\hypersetup{
    colorlinks,
    linkcolor={blue!80!black},
    citecolor={blue!80!black},
}
\tikzset{
    root/.style =             {align=center, text width=1cm, rounded corners=3pt, line width=0.3mm, fill=gray!10, draw=gray!80, font=\small},
    demographic/.style =         {align=center, text width=1.8cm, rounded corners=3pt, line width=0.3mm, fill=blue!10, draw=blue!80, font=\footnotesize},
    demographic_work/.style =    {align=center, text width=10cm, rounded corners=3pt, line width=0.3mm, fill=blue!10, draw=blue!0, font=\footnotesize},
    character/.style =         {align=center, text width=1.8cm, rounded corners=3pt, line width=0.3mm, fill=red!10, draw=red!80, font=\footnotesize},
    character_work/.style =    {align=center, text width=10cm, rounded corners=3pt, line width=0.3mm, fill=red!10, draw=red!0, font=\footnotesize},
    personalization/.style =           {align=center, text width=1.8cm, rounded corners=3pt, line width=0.3mm, fill=cyan!10, draw=cyan!80, font=\footnotesize},
    personalization_work/.style =      {align=center, text width=10cm, rounded corners=3pt, line width=0.3mm, fill=cyan!10, draw=cyan!0, font=\footnotesize},
    risk/.style =         {align=center, text width=1.8cm, rounded corners=3pt, line width=0.3mm, fill=orange!10, draw=orange!80, font=\footnotesize},
    risk_work/.style =    {align=center, text width=10cm, rounded corners=3pt, line width=0.3mm, fill=orange!10, draw=orange!0, font=\footnotesize},
}

\newcommand{\cmark}{\ding{51}}
\newcommand{\xmark}{\ding{55}}

\usepackage{CJK}

\def\vtheta{{\bm{\theta}}}
\def\vphi{{\bm{\phi}}}
\def\mP{{\bm{P}}}
\def\mX{{\bm{X}}}
\DeclareMathOperator*{\argmax}{arg\,max}

\newcommand{\method}{\textsc{AMix-1}\xspace}

\newcommand{\methodt}{\textsc{AMix-1}}
\newcommand{\ttsmethod}{\textsc{EvoAMix-1} }
\newcommand{\ttsmethodt}{\textsc{EvoAMix-1}}

\title{
    \methodt: A Pathway to Test-Time Scalable \\ Protein Foundation Model
}

\renewcommand{\affiliationlist}{%
\vspace{-5mm}
  \begingroup\affiliationfont\fontfamily{soramedium}\selectfont
  $^{1}$Shanghai Artificial Intelligence Laboratory\\[1pt]
  $^{2}$Generative Symbolic Intelligence Lab (GenSI), Tsinghua University\\[1pt]
  $^{3}$Institute for AI Industry Research (AIR), Tsinghua University\\[2pt]
  $^{4}$Tsinghua University\enspace
  $^{5}$Fudan University\enspace\\
  $^{6}$Tianjin University\enspace
  $^{7}$Georgia Institute of Technology\\[1pt]
  $^{8}$Beijing University of Posts and Telecommunications\enspace\\
  $^{9}$University of Chinese Academy of Sciences\enspace
  $^{10}$City University of Hong Kong%
  \endgroup
  \vspace{-4mm}
}

\abstract{

We introduce \method, a powerful protein foundation model built on Bayesian Flow Networks and empowered by a systematic training methodology, encompassing pretraining scaling laws, emergent capability analysis, in-context learning mechanism, and test-time scaling algorithm. 
To guarantee robust scalability, we establish a predictive scaling law and reveal the progressive emergence of structural understanding via loss perspective, culminating in a strong 1.7-billion model. 
Building on this foundation, we devise a multiple sequence alignment~(MSA)-based in-context learning strategy to unify protein design into a general framework, where \method recognizes deep evolutionary signals among MSAs and consistently generates structurally and functionally coherent proteins. 
This framework enables the successful design of a dramatically improved AmeR variant with an up to $50\times$ activity increase over its wild type. 
Pushing the boundaries of protein engineering, we further empower \method with an evolutionary test-time scaling algorithm for \textit{in silico} directed evolution that delivers substantial, scalable performance gains as verification budgets are intensified, laying the groundwork for next-generation lab-in-the-loop protein design.

    \vspace{-2.5mm}
}

\correspondence{\href{mailto:zhouhao@air.tsinghua.edu.cn}{zhouhao@air.tsinghua.edu.cn}}

\projectpage{\url{https://gensi-thuair.github.io/AMix-1/}}

\begin{document}

    \maketitle
    
    \vspace{-16mm}

    \begin{figure}[H]
        \centering
        \includegraphics[width=0.73\linewidth]{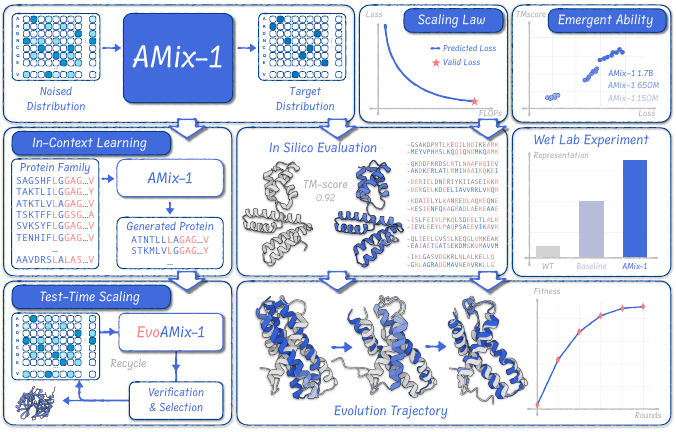}
        \caption{An overview of \method. \method is built upon a pathway methodology of scaling law, emergent ability, in-context learning, and test-time scaling, and is validated on both \textit{in silico} evaluation and wet-lab assay.}
        \label{fig:overview}
    \end{figure}

    \newpage
    \tableofcontents
    \newpage

    \section{Introduction}

Advanced artificial intelligence models, such as AlphaFold~\cite{senior2020improved,varadi2022alphafold,abramson2024accurate} and ESM~\cite{rives2021biological,lin2023evolutionary,hayes2024simulating}, have significantly facilitated a wide range of protein research, including structure prediction~\cite{senior2020improved,varadi2022alphafold,abramson2024accurate,lin2022language,wu2022high,fang2023helix}, inverse folding~\cite{hsu22learning}, functional property prediction~\cite{jang2024accurate,yuan2024genome}, mutational effect estimation~\cite{notin2022tranception,su2023saprot,glaser2025esmeffect}, and \textit{de novo} protein design~\cite{hayes2024simulating,ni2024forcegen}.

However, these models lack a unified, scalable, and systematic methodology akin to that of cutting-edge large language models (LLMs)~\cite{brown2020language,zeng2023glm130b,openai2024gpt4,deepseekai2025deepseekv3,deepseekai2025deepseekr1,yang2025qwen3,geminiteam2025gemini}, whose capabilities consistently improve with increased scale in data, model size, and compute.
Consequently, the systematic trading of compute for performance has revealed unprecedented emergent intelligence, as models continue to improve with larger computational investments, both during pretraining~\cite{hestness2017deep, rae2022scaling, hoffmann2022training} and at inference~\cite{snell2024scaling, brown2024large, wu2025inference}.
This observation motivates a fundamental question:
\begin{DefinitionBox}
As large language models begin to exhibit traits of artificial general intelligence, how can we craft protein foundation models under a converging methodology grounded in \textbf{scaling law} characterization, \textbf{emergent capability} analysis, \textbf{in-context learning} mechanism, and \textbf{test-time scaling} algorithms, to unlock scalable and universal protein design of high fidelity?
\end{DefinitionBox}

In this study, we propose \method, a pathway methodology for scalable protein design~(\Cref{fig:overview}).
We first characterize a predictable scaling law for Bayesian Flow Networks~\cite{graves2023bayesian} and reveal that structural understanding emerges naturally as training progresses, with which we pretrained a powerful 1.7B \method.
Then we leverage its in-context learning capability with multiple sequence alignments to unify structure- and function-guided protein design into a general framework.
With \method, we successfuly design a novel AmeR variant validated through wet-lab experiments, achieving a $50\times$ activity increase over the wild type.
To push the limits further, we fledge \method with an evolutionary test-time scaling (TTS) algorithm for \textit{in silico} directed evolution, allowing plug-and-play metrics and delivering strong performance gains with increasing verification budgets.

We summarize the main contributions of our study as follows:
\begin{itemize}
\item We propose a systematic paradigm, including scaling laws, structural emergent ability, in-context learning, and test-time scaling, for crafting protein foundation models, emphasizing a road map towards scalable protein design;
\item We are the first to characterize scaling law and emergent ability for an advanced generative model, i.e., Bayesian Flow Network, and pretrain a family of powerful models, \methodt, scaling from 8 million to 1.7 billion parameters;
\item We introduce a novel evolutionary test-time scaling algorithm by leveraging \methodt's in-context learning for unified protein design, amplifying its effectiveness in \textit{in silico} directed evolution, delivering rapid gains as verification budgets grow;
\item We design a high-activity AmeR variant with a $50\times$ improvement over the wild type, showcasing \methodt’s real-world impact on functional protein engineering.
\end{itemize}

    \section{Protein Modeling with Bayesian Flow Network}
\label{sec:bfn}
Protein modeling with bayesian flow networks (BFNs)~\cite{graves2023bayesian} introduces a generative framework that learns the continuous parameters of protein sequence distributions through iterative Bayesian updates, rather than directly modeling discrete amino acid sequences. This process iteratively refines a set of noisy samples initiated from an uninformative prior $\vtheta_0$, producing successive posteriors $\vtheta_i$ that gradually incorporate more information and increase certainty.

The generative process is modeled as a message-passing protocol between \emph{sender distribution} and \emph{receiver distribution}, where the sender's visibility is confined to the sample space, while the receiver infers messages based on its interpretation of samples and parameters. 

During each communication round, the \emph{sender distribution} $p_\text{s}(\mathbf{y} \mid \mathbf{x}; \alpha)$ perturbs protein sequences $\mathbf{x}$ with a noise scheduler $\alpha$ and sends the noisy protein sequences $\mathbf{y}$ to the receiver distribution, which mirrors a forward diffusion mechanism. Specifically, for the protein sequence distribution, the sender distribution can be formulated as follows:
\begin{equation}
p_\text{s}(\mathbf{y} \mid \mathbf{x}; \alpha) = \mathcal{N}\left( \mathbf{y} \mid \alpha (K \mathbf{e}(\mathbf{x}) - \mathbf{1}), \alpha K \mathbf{I} \right)
\end{equation}
where $\mathbf{e}(\mathbf{x}) $ is the one-hot encoding of the protein sequence $\mathbf{x}$, flattened into a vector, with each position encoded as a $K$-dimensional vector, and $K$ is the number of amino acid types. Here $\alpha$ is derived from the schedule $\beta(t)$, where $\beta_1 \in R^+$ is the final noise level at $t = 1$:
\begin{equation}
\alpha = \beta(t) = \beta_1t^2
\end{equation}
On the other hand, the \emph{receiver distribution} employs a neural network $\vphi$ to approximate the sender distribution with its current belief $\vtheta_{i-1}$. Concretely, the network $\vphi$ takes the belief state $\vtheta_{i-1}$ as input, and the corresponding distribution $p_\text{in}(\cdot|\vtheta)$ is termed the \emph{input distribution}. The network output $\vphi(\vtheta_{i-1})$ is an updated distribution $p_\text{out}(\cdot|\vphi(\vtheta_{i-1}))$ over the sample space, termed the \emph{output distribution}. Integrating the output distribution with the noise scheduler $\alpha$ yields the receiver distribution as follows:
\begin{equation}
p_\text{r}(\mathbf{y}_i\mid \vtheta_{i-1},\vphi,\alpha_i)= \mathbb{E}_{p_\text{out}(\mathbf{x}' |\vphi(\boldsymbol{\vtheta}_{i-1}))}p_\text{s}\left(\mathbf{y}_i \mid \mathbf{x}' ; \alpha_i\right)
\end{equation}
\begin{figure}[]
\centering
\includegraphics[width=0.98\linewidth]{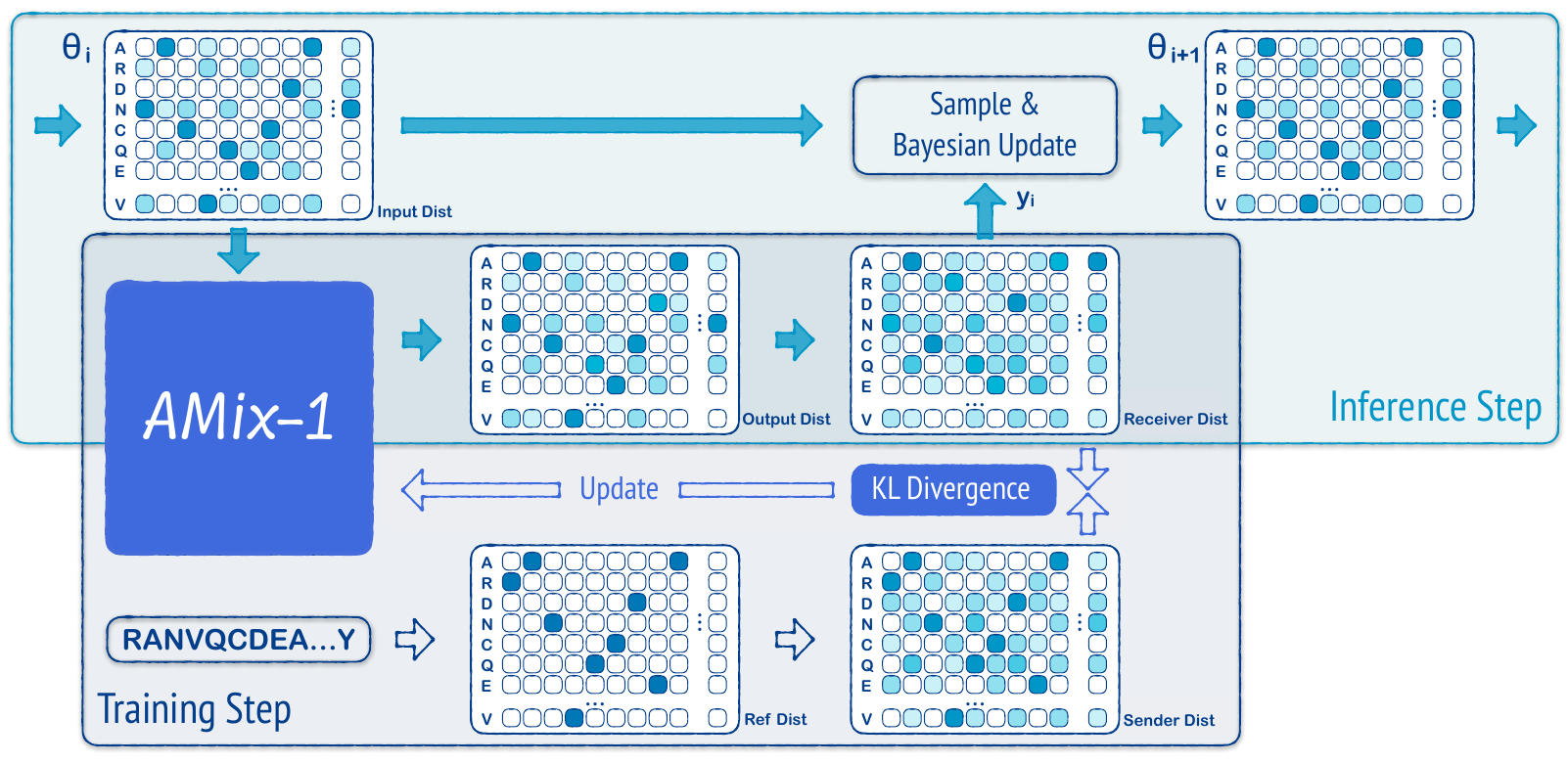}
\caption{
Illustration of Bayesian Flow Networks (BFNs).
The network $\vphi$ takes as input the distribution $p_{\text{in}}$ and generates the output distribution $p_{\text{out}}$.
The training objective is to minimize the Kullback-Leibler (KL) divergence between the sender distribution $p_s$ and the receiver distribution $p_r$, both of which are perturbed by additive noise $\alpha_i$.
}
\label{fig:bfn}
\end{figure}
Therefore, the final training objective integrates the KL divergence between sender and receiver distributions across timesteps and trajectories, yielding the following loss:
\begin{align}\label{equ:kl}
\mathcal{L}(\vphi) = \mathbb{E}_{\mathbf{x} \sim p_{\text{data}}}\mathbb{E}_{\Pi_{i=1}^n p_\text{s}(\mathbf{y}_i|\mathbf{x},\alpha_i)}  D_{KL} \left(p_\text{s}(\mathbf{y}_i|\mathbf{x},\alpha_i)||p_\text{r}(\mathbf{y}_i\mid \vtheta_{i-1},\vphi,\alpha_i)\right),
\end{align}
Leveraging the Bayesian update function to model the information gathering process, which iteratively refines priors into data-informed posteriors as follows:
 \begin{equation}
 \vtheta_i = h(\vtheta_{i-1}, \mathbf{y}_i, \alpha_i) =  \frac{e^\mathbf{y }\vtheta_{i-1}}{\sum_{k=1}^K e^\mathbf{y_k} (\vtheta_{i-1})_k}
\end{equation}

In practice, we use the continuous categorical values of output distribution  $\vphi(\boldsymbol{\vtheta}_{i-1})$ without sampling to directly update $\vtheta$ bypassing the sampling of noisy data needed for Bayesian update, which is similar to the noise reduced sampling method used in MolCRAFT~\cite{qu2024molcraft}. The effective sampling method is described in appendix Algorithm \ref{alg:noise_reduced_sampling}.
We also provide \Cref{fig:bfn} to help the readers visually grasp the key concepts of BFNs.

    \section{Scaling Law}
\label{sec:scaling}

This section begins by detailing the pretraining setup, followed by an investigation of the scaling laws that govern \method models.
These laws characterize the relationships among key factors, including cross-entropy loss, model size, the number of training tokens, and computational cost measured in FLOPs.

\begin{DefinitionBox}
\textbf{Key Result} \ding{172}: \methodt's cross entropy is precisely predictable against compute, data quantity and model size.
\end{DefinitionBox}

\subsection{Pretraining}

\paragraph{Datasets} The scaling law exploration in the pre-training stage is conducted using the \textbf{UniRef50} dataset, derived from UniProtKB~\cite{varadi2022alphafold} and selected UniParc sequences through iterative clustering (UniProtKB$+$UniParc $\rightarrow$ UniRef100 $\rightarrow$ UniRef90 $\rightarrow$ UniRef50)~\cite{suzek2007uniref,suzek2015uniref}.
This process ensures that UniRef50's representative sequences are high-quality, non-redundant, and diverse, providing broad coverage of the protein sequence space for protein language models.
Specifically, we utilized the pre-processed UniRef50 dataset provided by EvoDiff~\cite{alamdari2023protein}, comprising $41,546,293$ sequences for training and $82,929$ sequences for validation.
Sequences longer than $1,024$ residues were randomly cropped to $1,024$ to reduce computational costs and generate diverse subsequences.

\paragraph{Architecture}

\begin{table}[b]
\caption{Architecture of \method \;at various scales.}
\label{tab:archi}
\centering
\begin{tabular}{c|cccccc}
\toprule
Model Size & $8$M & $35$M & $150$M & $350$M & $650$M & $1.7$B \\ \hline
Number of layers & $6$ & $12$ & $30$ & $33$ & $33$ & $48$ \\
Hidden dimensions & $320$ & $480$ & $640$ & $960$ & $1280$ & $1680$ \\
Attention heads& $20$ & $20$ & $20$ & $20$ & $20$ & $40$ \\
\bottomrule
\end{tabular}
\end{table}
\method model series are built upon the encoder-only Transformer architecture~\cite{vaswani2017attention}.
To encode positional information while preserving sequence invariance in attention patterns, we incorporate Rotary Position Embeddings (RoPE)~\cite{su2024roformer}, which introduce relative position dependencies through a rotation-based mechanism applied to query and key vectors.
To explore the impact of model capacity on performance, we scale the architecture by systematically modifying the number of encoder layers, the number of attention heads, and the dimensionality of the hidden representations.
These architectural configurations and corresponding parameter counts are summarized in \Cref{tab:archi}.

\paragraph{Training Details}

All models were trained using the Adamw~\cite{loshchilov2018decoupled} optimizer with a learning rate of $4\times10^{-4}$, $\alpha$ values of $(0.9, 0.98)$, and a weight decay of $0.01$.
A polynomial learning rate schedule was adopted, spanning a total of $1$ million training steps with a warmup phase of $2,000$ steps.
During warmup, the learning rate increased linearly from $1\times10^{-7}$ to $4\times10^{-4}$, followed by a polynomial decay to $4\times10^{-5}$ with a power factor of $1$.
Training was performed using mixed-precision (BF16) and a distributed data parallel (DDP) setup.
To mitigate exploding gradients, gradient norms were clipped to a maximum of $1.0$.
Early stopping was applied based on validation loss, with a patience threshold of $1,000$ steps. 
For \method-1.7B and \method-650M, training was conducted with a batch size of $4$ million across 16 nodes, each equipped with $8$ A$800$ GPUs.
Smaller models were trained with a batch size of $1$ million.

\paragraph{Noise Level}
In BFN models, the noise level $\alpha$ is defined as: $\alpha=\beta(t) = \beta_1 \cdot t^{2}, (1\geq t \geq 0)$, where $\beta_1$ controls the initial noise magnitude.
Note that a smaller $\alpha$ corresponds to a higher noise intensity.
Unless otherwise specified, all \method models were trained with $\beta_1 = 1.0$.

\subsection{Predictive Scaling Behavior}\label{sec:data_law}

\paragraph{Motivations}

First, predicting the loss trajectory of a model \cite{chowdhery2023palm,chung2024scaling} has gained significant attention in the LLMs community.
The primary motivation lies in the substantial computational and financial costs of training LLMs.
Accurately forecasting final performance early in the training process enables practitioners to make informed decisions about whether to continue, halt, or adjust the training strategy.
This predictable capability mitigates the risk of wasted resources and enables more efficient exploration of architectural and training choices.
This motivation is particularly relevant in the context of BFN models, which also operate at scale and under varying noise levels.

Second, predictable loss modeling is closely linked to the phenomenon of emergence (see \Cref{sec:emergent}), in which specific downstream capabilities arise abruptly once the training loss falls below a critical threshold.
Consequently, forecasting the loss trajectory not only enhances computational efficiency but also offers a principled approach to anticipating the onset of emergent behaviors.

\paragraph{Experiment Settings and Metrics}\label{sec:scaling_setting}
Prior studies \cite{liang2024scaling,ravishankar2025scaling} often adopt an averaged loss across different noise levels to characterize the scaling behavior of diffusion-style models.
In contrast, for \method models, we investigate fine-grained scaling behavior under varying noise levels, rather than aggregating results over a single average noise configuration.
We make this decision for two main reasons.
First, BFNs naturally involve multi-objective learning across noise levels, making simple averaging inappropriate as it masks the distinct learning dynamics of each subtask.
Second, under extreme noise settings, BFN models often fail to converge, rendering them unscalable; averaging in such cases introduces excessive variance and distorts the scaling law.
Therefore, we compute the scaling relationship separately for each noise level $\alpha$ to capture a more faithful and fine-grained behavior of BFN models.

As for the metric, although we employ KL divergence (i.e., \Cref{equ:kl}) as the training objective, we report the cross-entropy between the output distribution $p_\text{out}$ and the ground-truth sequence, $\mathcal{L} = -\log p_\text{out}$, to evaluate scaling behavior.
This choice is motivated by two factors.
First, from an information-theoretic perspective, the cross-entropy measures the inefficiency of encoding the ground-truth sequence using the model’s predicted distribution.
Since the ground truth is inherently one-hot, minimizing cross-entropy directly reflects the model’s ability to recover precise, token-level information.
Second, prior works on scaling laws~\cite{rae2022scaling, hoffmann2022training, achiam2023gpt} have consistently adopted cross-entropy, allowing for direct and meaningful comparisons.

\begin{figure}[]
\centering
\includegraphics[width=0.98\linewidth]{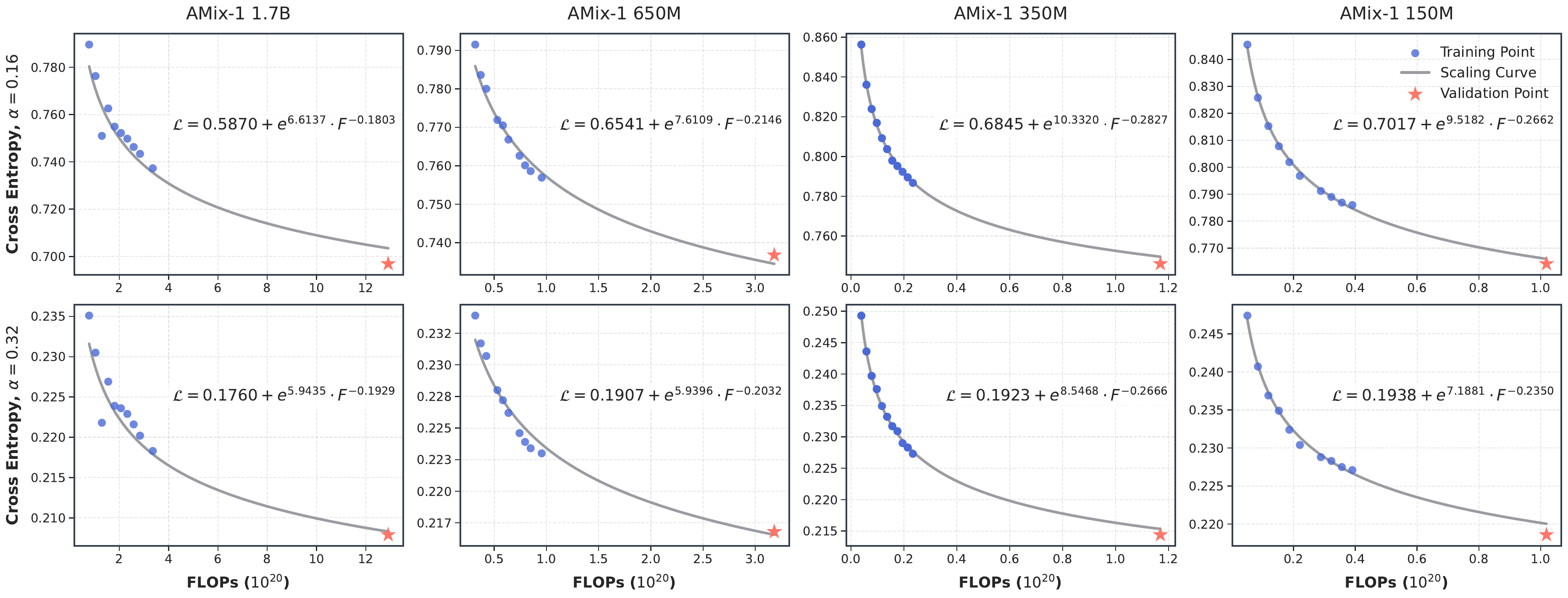}
\caption{Scaling laws of cross-entropy and training FLOPs across various model sizes and the noise level $\alpha$. We show the equation between $\mathcal{L}$ and FLOPs $F$, indicating that the \method models have predictive scaling laws.}
\label{fig:scaling}
\end{figure}

\paragraph{Scaling Law of Compute}
We begin by varying the number of training steps for our \method models, covering model sizes from 8M to 1.7B parameters.
Since the total training FLOPs $F$ can be approximated as $F \approx 6ND$, where $N$ denotes the number of model parameters and $D$ is the number of training tokens, fixing the model size allows us to use FLOPs as a proxy for training compute.
Specifically, for a given noise level $\alpha$, we examine the power-law relationship between cross-entropy loss $\mathcal{L}$ and training FLOPs $F$:
\begin{equation}\label{equ:flops_law}
\mathcal{L}(F) = E_0+e^a\cdot F^{b},
\end{equation}
where $E_0$ denotes the irreducible loss, representing the asymptotic lower bound that cannot be reduced through additional compute.
As shown in \Cref{fig:scaling}, our \method models exhibit clear scaling behavior under two noise levels $\alpha=0.16$ and $0.32$: the validated data points (red stars) align closely with the extrapolated power-law curves, consistent with the scaling laws observed in large language models~\cite{achiam2023gpt,openai2024gpt4}.
This consistency enables reliable performance forecasting and informs efficient compute allocation for larger training regimes.

In addition to moderate noise levels ($\alpha = 0.16$ and $0.32$), we evaluate $\mathcal{L}$ under extreme noise conditions ($\alpha = 0.08$ and $0.64$), as shown in \Cref{fig:unscaling}.
Under these settings, the BFN models fail to capture the training objective defined in \Cref{equ:kl}, leading to erratic and unpredictable cross-entropy values.
What's more, although both $\alpha=0.16$ and $\alpha=0.32$ yield similar scaling trends, the subtle differences in their exponents indicate that careful scheduling of $\alpha$ can significantly influence convergence behavior and generalization.
Since $\alpha$ directly modulates the injected noise, we point out that \textbf{an appropriate schedule of $\alpha$ when training may be essential for stabilizing training and achieving optimal performance in BFN-based models}.

\begin{table}[b]
\centering
\caption{Different parametric scaling laws under various noise levels. Note that both $N$ and $D$ are measured in millions.}
\resizebox{0.8\linewidth}{!}{
\begin{tabular}{c| c c c c c | c c | c}
\toprule
\multicolumn{1}{c}{\textbf{Noise Level}} & \multicolumn{5}{c|}{\textbf{Parameters}} & \multicolumn{2}{c|}{\textbf{MRE (\%)}} & \multirow{2}{*}{\textbf{Scalable}} \\
\cmidrule(lr){1-1} \cmidrule(lr){2-6} \cmidrule(lr){7-8}
$\alpha$ & $E$ & $A$ & $B$ & $n$ & $d$ & Fitting & Valid & \\
\midrule
$0.02$ & $2.5941$ & \textcolor{red!90!black}{0.0000} & $0.7461$ & $0.2595$ & $0.4891$ & $0.56$ & $0.55$ & \textcolor[HTML]{FF7E79}{\xmark}\\ 
\midrule
$0.16$ & $0.4638$ & $0.2570$ & $4.6740$ & $0.0380$ & $0.3332$ & $0.73$ & $1.66$ & \textcolor[HTML]{4A69D6}{\cmark}\\ 
\midrule
$0.32$ & $0.1703$ & $0.0122$ & $0.5813$ & $0.4850$ & $0.2047$ & $1.06$ & $1.57$ & \textcolor[HTML]{4A69D6}{\cmark}\\ 
\midrule
$0.64$ & $0.0278$  & \textcolor{red!90!black}{0.0000} & $0.5758$ & $0.3088$ & $0.5887$ & $2.22$ & $5.92$ & \textcolor[HTML]{FF7E79}{\xmark}\\ 
\bottomrule
\end{tabular}
}
\label{tab:paramscaling}
\end{table}

\subsection{Optimal Estimation under Fixed FLOPs}

In this section, we address the following question:
Given a fixed training FLOPs budget, how should it be optimally allocated between model size and the number of training tokens for \method models?
Following the methodology of Chinchilla \cite{hoffmann2022training}, we examine this trade-off from two complementary perspectives.

\paragraph{Parameter-aware Scaling Dynamics}
Under a fixed compute budget, it is important to determine an optimal balance between model size and data scale.
To this end, we adopt a parametric scaling law that models the cross-entropy $\mathcal{L}$ as a joint function of the number of model parameters $N$ and the number of training tokens $D$.
Specifically, under a fixed noise level, we fit the cross-entropy across multiple configurations of $(N, D)$ using the following formulation:
\begin{equation}\label{equ:parametric_law}
\mathcal{L}(N,D) = E + \frac{A}{N^{n}} + \frac{B}{D^{d}},
\end{equation}
where $E$ represents the irreducible loss floor, and the two additive terms quantify the diminishing returns from increasing model size and training data, respectively.

As discussed in \Cref{sec:scaling_setting}, we investigate the parametric scaling law of \method models across different noise levels $\alpha$, with detailed results summarized in \Cref{tab:paramscaling}.
Additional fitting results for other noise settings are provided in \Cref{app:param_scale}.
To ensure robustness and assess generalization, we split the data by training tokens, using the former $30\%$ for fitting and the latter $30\%$ for validation.
Following \citet{hoffmann2022training}, we minimize the Huber loss \cite{huber1992robust} between the
predicted and observed log loss using the L-BFGS algorithm \cite{nocedal1980updating} to estimate \Cref{equ:parametric_law}.
We quantify the goodness of fit using two metrics: the Fitting Mean Relative Error (MRE) and the Validation MRE, defined in \Cref{app:mre}.
These metrics reflect the relative prediction accuracy within and beyond the fitting region, respectively.

Similar to the results in the scaling law of compute, the parametric scaling law is also $\alpha$-aware.
For example, $\alpha=0.02$ and $0.64$ exhibit unscalable behaviors, as evidenced by the parameter $A$ being effectively estimated as $0.000$.
This implies that under these settings, increasing the number of model parameters $N$ brings no improvement to the cross entropy $\mathcal{L}$.
The vanishing $A$ values at both ends of the $\alpha$ spectrum suggest that only within a moderate noise regime can model size effectively trade-off against training tokens to reduce the cross-entropy.
This is precisely what we observe at $\alpha=0.16$ and $0.32$, where $A$ is significantly non-zero, and both Fitting and Valid MREs are small, confirming the presence of a predictive, scalable relationship.

\paragraph{Compute-Optimal Frontier}

\begin{figure}[t]
\centering
\begin{subfigure}[]{0.49\linewidth}
\includegraphics[width=\linewidth]{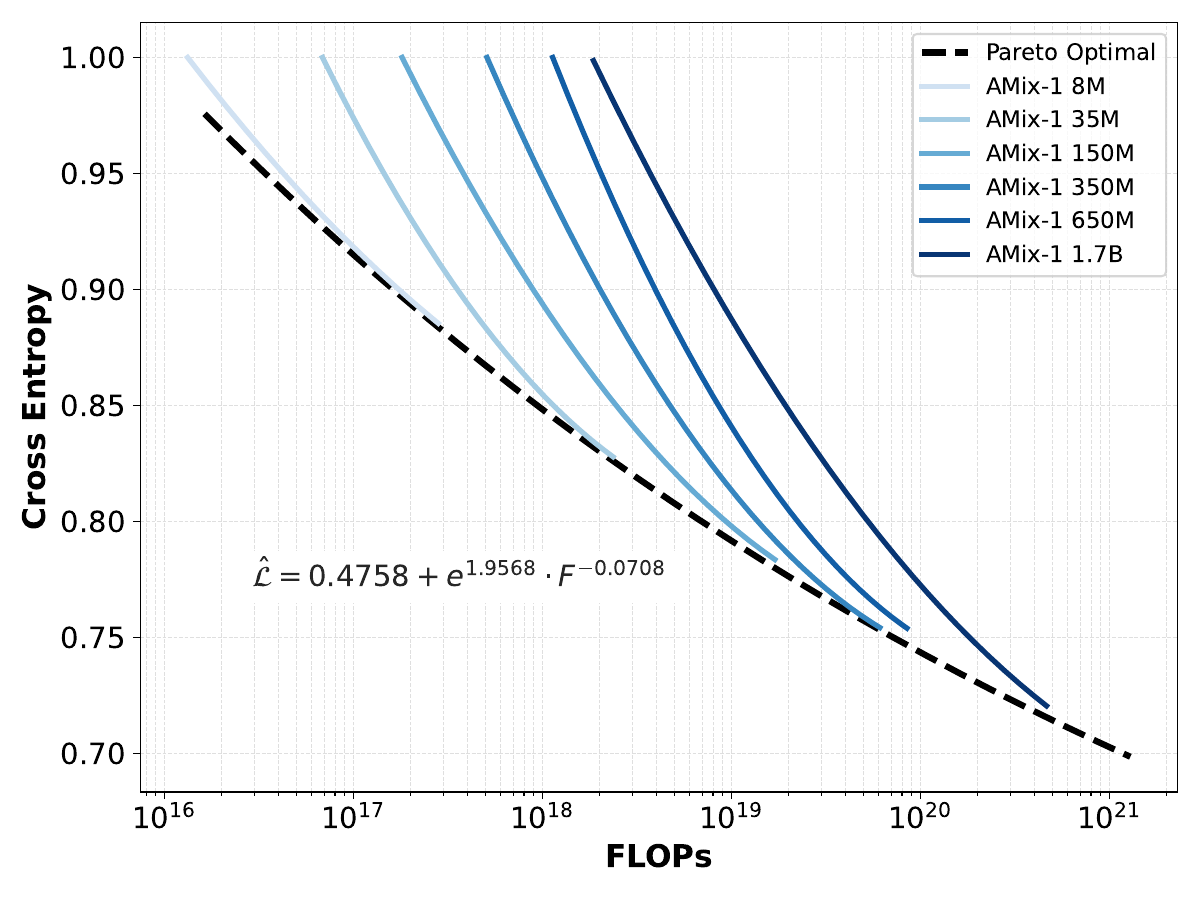}
\caption{Pareto frontier when $\alpha = 0.16$}
\label{fig:pareto016}
\end{subfigure}
\hfill
\begin{subfigure}[]{0.49\linewidth}
\includegraphics[width=\linewidth]{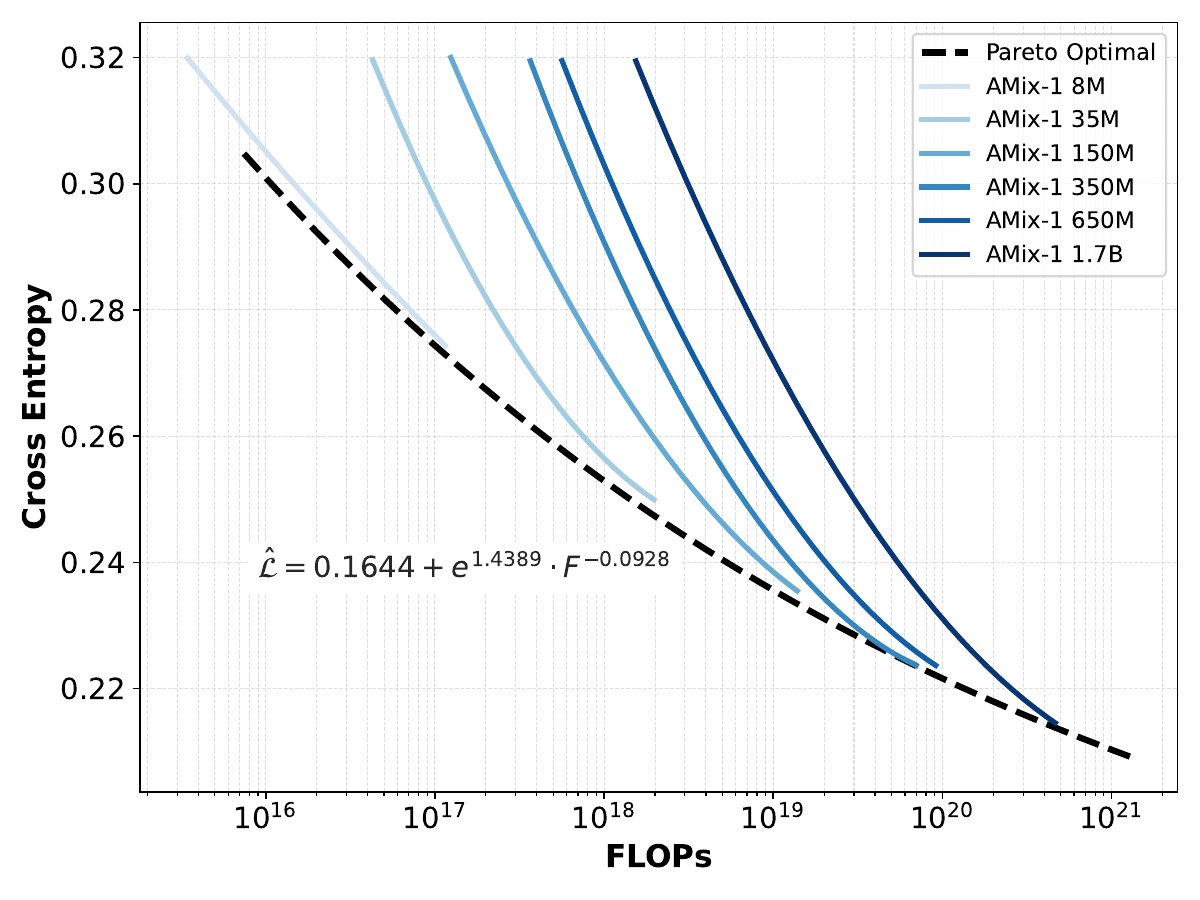}
\caption{Pareto frontier when $\alpha = 0.32$}
\label{fig:pareto032}
\end{subfigure}
\caption{Pareto optimal curves comparing validation loss versus FLOPs for six different model scales. 
The dashed black lines denote the Pareto frontiers, capturing the best achievable trade-offs between compute and validation loss under different noise levels.}
\label{fig:pareto}
\vspace{-3mm}
\end{figure}

Summarizing the above analysis, we conclude that as the compute budget increases, both model size and the number of training tokens should scale approximately proportionally to maintain optimal performance.
Following this principle, we adopt the approach of \citet{hoffmann2022training} to estimate a \textbf{compute-optimal frontier}, which serves as a guideline for jointly managing the model size and data scale under a fixed FLOPs budget.

For each noise level $\alpha$ under which BFN models exhibit scalable behavior, we analyze the compute-optimal Pareto frontier to identify the most efficient combination of model size and training token count that minimizes validation loss under a fixed computational budget.
This frontier delineates the best attainable trade-off between parameter count and data usage, offering a principled framework for resource allocation across varying noise regimes.
Under a given noise level $\alpha$, our objective is to characterize the compute-optimal frontier that relates the minimum achievable cross-entropy $\hat{\mathcal{L}}$ to the training FLOPs $F$, similar to \Cref{equ:flops_law}:
\begin{equation}
\hat{\mathcal{L}}=E_0+e^{a}\cdot F^{-b}.
\end{equation}
To accurately estimate the scaling parameters $a$ and $b$, we analyze the cross-entropy loss as a function of FLOPs for six \method model variants ranging from 8M to 1.7B parameters.
In \Cref{fig:pareto}, we present two compute-optimal frontiers for $\alpha = 0.16$ and $\alpha = 0.32$.
Each curve corresponds to a fixed model size, with dots denoting empirical measurements under different training regimes and solid lines representing smoothed scaling trends.
Larger models such as \method-650M and \method-1.7B display steeper loss reductions, suggesting greater efficiency in leveraging additional compute.
In contrast, smaller models, such as \method-8M and \method-35M, saturate earlier, indicating a limited capacity to benefit from extended training.
The dashed black line shows the global Pareto frontier, fit across all configurations.
Notably, \method-650M and \method-1.7B closely align with this frontier, demonstrating near-optimal compute efficiency.
Conversely, the pronounced deviation of smaller models from the frontier highlights their sub-optimality in the compute–performance trade-off.

    \section{Emergent Ability}
\label{sec:emergent}

Emergent ability~\cite{wei2022emergent} is a critical topic in large language modeling, as it seeks to understand how high-level capabilities arise, such as reasoning, as foundation models become more powerful with increasing size and computational investment. 
In the context of protein foundation models, studying emergent abilities is equally important for understanding how the capacity to represent and generate protein sequences, structures, and even functions evolves as the training objective is progressively optimized. 
Such investigations help determine whether a specific modeling goal is achievable by continually scaling data quantity, model size, and compute within a given foundation model framework.

Building on the predictive scaling law established in \Cref{sec:scaling}, we further investigate the less predictive aspects of \method's performance in protein generation. 
Our analysis of emergent behavior adopts a loss-centered perspective, following~\citet{du2024understanding}.
We adopt the predictive cross-entropy loss in \Cref{sec:scaling} as an anchor to empirically map training loss to protein generation performance. 
This empirical mapping provides a practical framework for interpreting how emergent capabilities arise during training, offering insight into the relationship between optimization dynamics and functional outcomes in protein modeling.

\begin{DefinitionBox}
\textbf{Key Result} \ding{173}: During the training process of \method, its structural understanding progressively emerges, with a consistent empirical alignment to its cross-entropy across model capacity.
\end{DefinitionBox}

\subsection{Experimental Setup}

In protein sequence learning, a central question is how structural information emerges as sequences are progressively learned, an inquiry grounded in the widely accepted ``sequence-structure-function'' paradigm~\cite{anfinsen1973principles}. 
To investigate this emergent process, we focus on three key capabilities: \textbf{sequence consistency}, reflecting sequence-level recovery from corrupted sequence distribution; \textbf{foldability}, representing the transition from sequence understanding to structural feasibility; and \textbf{structural consistency}, indicating the model's ability to preserve structural features. 
Notably, we demonstrate that structural consistency plays a critical role, as it underpins the emergence of few-shot learning capabilities discussed in \Cref{sec:few-shot}.

\paragraph{Sequence Consistency}
We evaluate the sequence consistency of the generated protein sequences $\mathbf{y} = \{y_1, \ldots, y_m\}$ by measuring their similarity to the corresponding input sequences $\mathbf{x} = \{x_1, \ldots, x_m\}$, using the metric as $\frac{1}{m} \sum_j \mathbf{I}[x_j = y_j]$. This metric quantifies the model’s ability to recover a corrupted distribution at the sequence level, aligning with the self-supervised pretraining objective. A higher sequence consistency indicates that the generated sequence closely follows the corrupted input, suggesting stronger adherence to the original distribution, while a lower SC score implies greater novelty in the generated output.

\paragraph{Foldability} 
Foldability serves as a critical assessment bridging the gap between protein sequence and structure.
Following recent practices in the community, we evaluate the foldability of generated sequences using ESMFold~\cite{lin2023evolutionary}. 
For each generated sequence, we predict its three-dimensional structure and extract the corresponding predicted Local Distance Difference Test (pLDDT) score~\cite{jumper2021highly}. 
The pLDDT score acts as a proxy for local structural confidence, with higher values indicating more reliable and well-formed structural predictions. 
From a foldability standpoint, elevated pLDDT scores suggest that the generated sequences are more likely to adopt stable, well-defined conformations, rather than collapsing into disordered or non-folding structures. 
Thus, this metric offers a valuable indication of the biophysical plausibility of novel proteins generated by the model.

\paragraph{Structural Consistency}
To assess how well \method, trained with a sequence-level objective, preserves structural consistency between generated sequences and their corresponding inputs, we employ the TM-score—a widely used metric for quantifying structural similarity between protein conformations. 
Specifically, we compare the three-dimensional structures predicted by ESMFold~\cite{lin2023evolutionary} for both the input and generated sequences. 
The TM-score ranges from 0 to 1, with higher values indicating greater structural alignment; scores above 0.5 typically suggest that the two structures share the same overall fold. 
From the perspective of sequence-conditioned generation, a high TM-score indicates that the model successfully retains the global structural scaffold of the input protein while introducing meaningful sequence-level diversity. 
This consistency metric is especially important in applications where structural conservation is essential due to functional or evolutionary constraints.

\begin{figure}[t]
\centering
\includegraphics[width=0.98\linewidth]{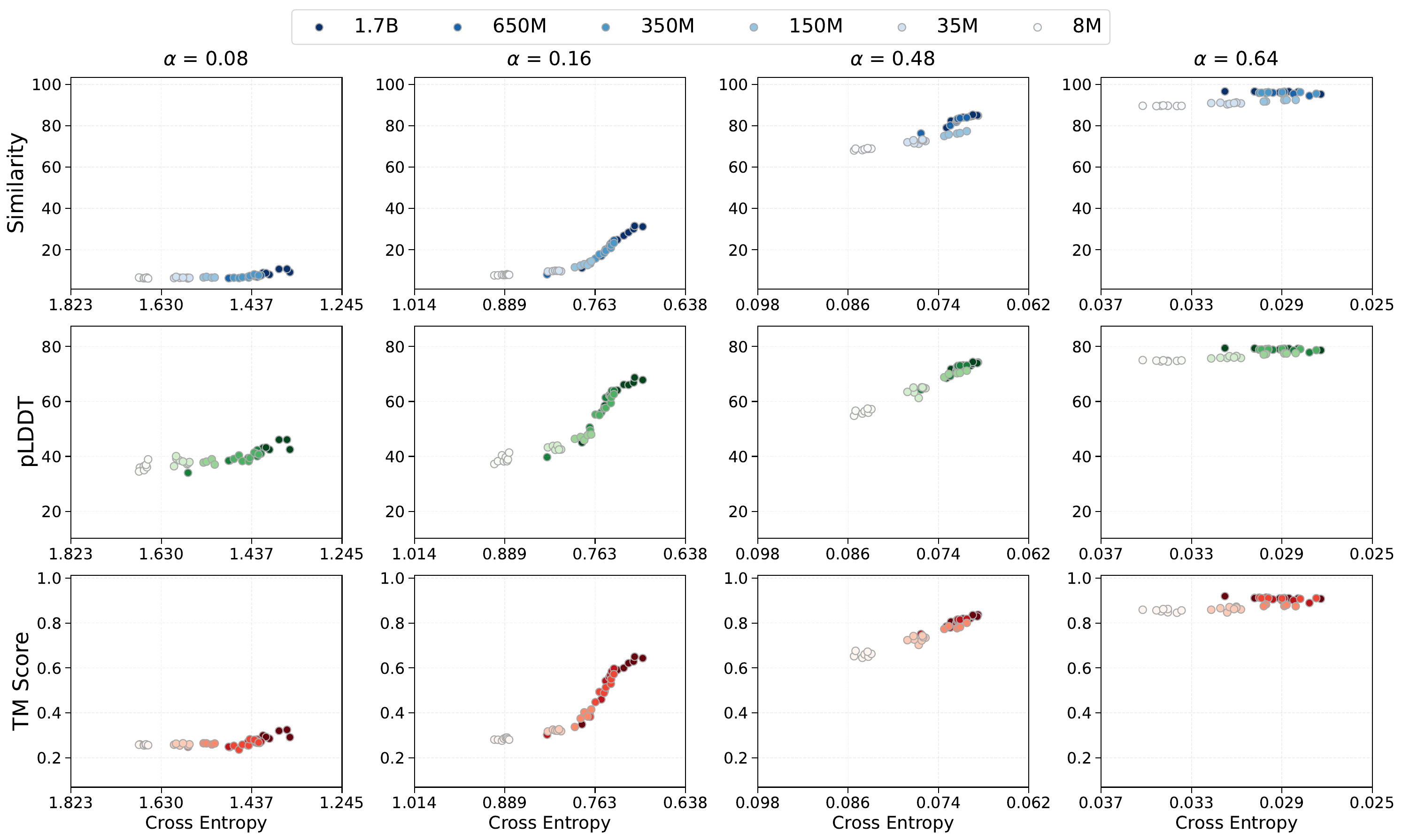}
\caption{Emergent abilities of \method models under varying noise levels $\alpha$. Noise levels of $\alpha=0.16$ and $0.48$ exhibit clear emergence. Darker colors indicate models with more parameters. Lower cross-entropy, and consequently stronger downstream performance, emerges as the model size increases.}
\label{fig:emergent}
\end{figure}

For the above metrics, we take Automated Model Evaluation (CAMEO)~\cite{robin2021continuous} as our evaluation benchmark. 
It is a continuously updated benchmark dataset that provides newly released, high-quality protein structures from the Protein Data Bank (PDB) for real-time assessment of protein structure prediction methods.
To assess the scalability of our models, we randomly selected $20$ proteins from CAMEO. Details of these sequences are provided in \Cref{app:cameo20}.

In this emergent ability analysis, \method adopts a setup consistent with the scaling law experiments described in ~\Cref{sec:scaling}. 
We do not apply additional decoding strategies, unlike previous foundation models~\cite{wang2024diffusion,atkinson2024protein}, in order to stabilize the emergent ability analysis and avoid introducing confounding effects.
We structure the analysis along the axis of corruption noise, denoted as $\alpha$. 
Specifically, we investigate four representative values: $\{0.08, 0.16, 0.48, 0.64\}$, where $\alpha = 0.08$ and $\alpha = 0.64$ correspond to extremely high and low noise levels, respectively, providing a comprehensive view of emergent behavior across the noise spectrum. 
These same $\alpha$ values are also used as the initial noise levels in \method's generation process. 
Our goal is to align the cross-entropy dynamics at these specific noise levels with the corresponding generation behavior, enabling a direct mapping between training metrics and generation performance.

\subsection{Findings}

We conduct emergent ability analysis using sequence consistency, foldability, and structural consistency as evaluation axes, and use the starting noise levels $\alpha \in \{0.08, 0.16, 0.48, 0.64\}$ as the noise dimension. The results are presented in \Cref{fig:emergent}, from which we draw several key findings summarized below.

\paragraph{Emergent capabilities are governed by the training objective.}
Model capabilities emerge as a direct consequence of optimizing the underlying training objective, underscoring the central role of loss minimization in the development of foundation models. 
Notably, we observe that, regardless of parameter count or training stage, all evaluated metrics are consistently aligned with the cross-entropy loss at each corruption level $\alpha$. 
These findings suggest that emergent abilities in BFN-based models are not solely a function of increasing model size; rather, they critically depend on the degree of epistemic uncertainty introduced via the noise parameter $\alpha$. 
This alignment between loss and capability offers a principled metric, as a ruler, for assessing the expected performance of newly trained models, as predicted by the established scaling law. 
Leveraging this ruler, we successfully train \method-1.7B, ensuring that it achieves the desired emergent behaviors with predictive outcomes.

\paragraph{Structural understanding emerges as the sequence objective is optimized.}
Despite being trained purely with sequence-level objectives, \method exhibits progressively stronger structural awareness as training advances, demonstrating that structural capabilities can indeed emerge without explicit structural supervision. 
We observe that as the sequence-level metric, namely cross-entropy, continues to decrease, both structural metrics (pLDDT and TM-score), along with sequence consistency, begin to improve rapidly once the cross-entropy drops below a certain threshold. 
Focusing on a representative case with median noise level $\alpha=0.16$, we find that performance initially plateaus: structural metrics show limited improvement even as the model continues to optimize its loss. 
However, upon further investment of compute, and as cross-entropy reaches a sufficiently low value, the model suddenly exhibits a sharp increase in structural performance—an effect commonly referred to as emergent ability. 
This observation is encouraging, as it suggests that structural understanding can emerge reliably from purely sequence-based, self-supervised learning, provided that sufficient optimization and compute are applied.

\paragraph{Emergent behavior is suppressed at large and small noise levels.}
Models trained at extreme noise levels, either too low or too high, fail to exhibit emergent capabilities, suggesting that only intermediate noise regimes provide sufficient signal for scalable generalization. 
Specifically, we observe that improvements in structural prediction metrics, including pLDDT, TM-score, and structural similarity, emerge under intermediate noise levels, particularly at $\alpha = 0.16$ and $\alpha = 0.48$. 
In these regimes, as cross-entropy loss decreases, \method consistently demonstrates substantial performance gains across all three metrics, regardless of model size. 
These improvements only materialize when the model is sufficiently trained to reduce cross-entropy below a critical threshold and when the injected noise balances exploration and convergence during training. 
Biologically, this shift reflects a transition from residue-level prediction to coherent, fold-level modeling, an indicator of structurally grounded intelligence.
 In contrast, no emergent behavior is observed at extreme values of $\alpha$, such as $\alpha = 0.08$ (high noise) or $\alpha = 0.64$ (low noise). 
At high noise levels, training dynamics become dominated by excessive stochasticity, which impairs the model’s ability to extract stable structural patterns. 
Despite continued optimization of the loss function, all structural performance metrics stagnate at low levels, showing no meaningful improvement with increased model size. 
Conversely, at low noise levels, the near-deterministic training dynamics lead to premature convergence and overfitting. 
The model fails to explore diverse structural configurations, limiting its capacity to generalize beyond the training data. 
In both cases, model scaling alone is insufficient to yield qualitative improvements in structure prediction. 
These findings underscore that emergent capabilities require not just increased scale but also a well-calibrated training noise level within an appropriate operating range.

\section{In-Context Learning}
\label{sec:few-shot}
\begin{figure}[tb]
    \centering
    \includegraphics[width=1\linewidth]{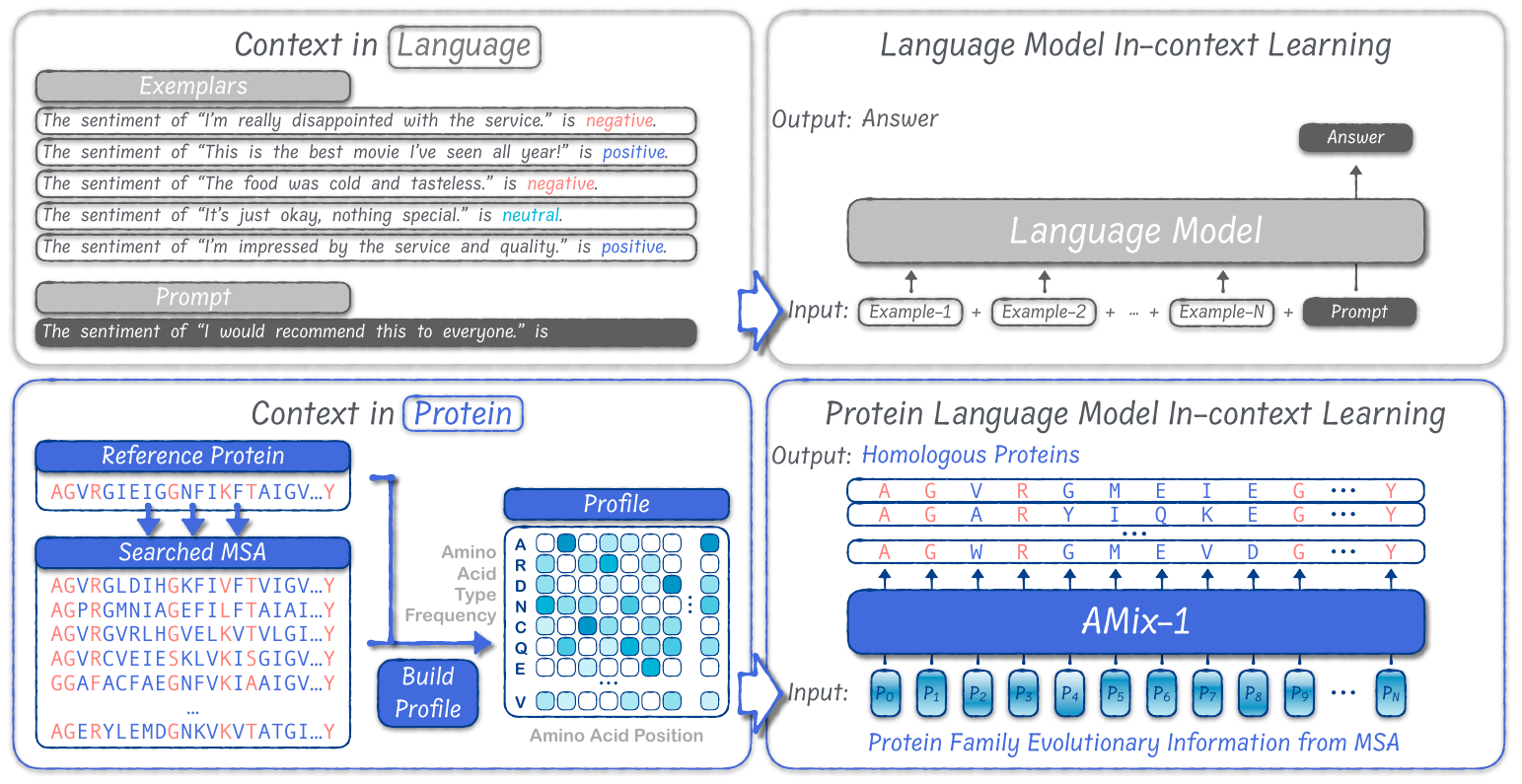}
    \caption{Comparison of standard LLM in-context learning (top) with protein family conditioning via MSA profiles in \method (bottom). In LLM ICL, few-shot exemplars are token-level demonstrations; in \method, the MSA profile serves as a dense, position-wise family prompt.}
    \label{fig:ICL}
\end{figure}

In-context learning (ICL) emerges as a central capability that enables large language models learns from a few contextual exemplars to perform unlimited tasks without task-specific fine-tuning~\cite{brown2020language,dong2022survey}.
This intriguing ability motivated us to develop a training-free strategy \textbf{to unify diverse protein design tasks by leveraging the in-context learning ability of \method}, which also serves as the foundation of our test-time scaling algorithm described in \Cref{sec:tts}.
In this section, we propose a novel approach that conditions \method on evolutionary profiles as prompts, enabling the generation of proteins that preserve both structural integrity and functional relevance.
We validated the effectiveness and generalizability of this approach through extensive \textit{in-silico} evaluations in structural and functional consistency.
Wet-lab experiments further confirmed its potential, successfully engineering the AmeR transcriptional repressor with 50-fold increase in activity over the wild type.

\begin{DefinitionBox}
\textbf{Key Result} \ding{174}: \method unifies diverse protein design tasks by leveraging its in-context learning ability to generate structure- and function-preserved proteins with a conditional protein profile.
\end{DefinitionBox}

\subsection{\method is a Unified Protein Designer}
\label{sec:msa}

We now introduce how evolutionary prompts are constructed to enable in-context learning in \method.
In the ICL framework of \method, we treat MSAs as prompts that guide generation toward a target protein family. 
MSAs encode evolutionary constraints—such as conserved motifs and co-evolving positions—which are critical for generating structurally and functionally valid sequences. 
Rather than conditioning directly on raw alignments, we transform the MSA into a position-wise frequency profile $\mP$, which serves as a dense, compressed representation of the family prompt.

Formally, given an alignment matrix $\mX \in \{0,\dots,K\}^{n \times m}$ of $n$ homologous sequences aligned to length $m$, we define the profile $\mP = \{\mP_i\}_{i=1}^{m}$, where each $\mP_i \in \Delta^{K-1}$ is a categorical distribution over amino acids at position $i$. 
Each component is computed as $\mP_{i,k} = \frac{1}{n} \sum_{j=1}^{n} \mathbb{I}[\mX_{ji} = k]$, where $k$ indexes the amino acid alphabet of size $K$, $\mP_{i,k}$ is the frequency of amino acid $k$ at position $i$ in the MSA, and $\mathbb{I}[\cdot]$ denotes the indicator function.

This transformation serves two purposes. First, it introduces a mild form of noise by averaging over observed residues—naturally aligning with \method’s denoising training objective. 
Second, by compressing MSAs of arbitrary depth into a fixed-size matrix, the profile representation makes \method invariant to the number of input sequences. 
This not only enables consistent inference efficiency regardless of MSA depth, but also simplifies the conditioning interface to the model~\cite{jingjing2025ProfileBFN}.

Given a profile prompt $\mP$, \method models the conditional probability over sequences in the family as:
\begin{equation}
p_{\theta}(\mathbf{x} \mid \mP) = \prod_{i=1}^{m} p_{\theta}(x_i \mid \mP)
\end{equation}
where $\mathbf{x} = (x_1, \dots, x_m)$ is the generated sequence and $\theta$ denotes the frozen model parameters. During decoding, we select the most probable residue at each position via $\argmax_k \, p_{\theta}(x_{i,k} \mid \mP)$.

In this way, \method acts as a conditional generator over the protein manifold, using the MSA-derived profile as a soft constraint that guides generation toward the desired family. This enables general-purpose, structure- and function-preserving design without any parameter updates.

\subsection{In Silico: \method Excels at In-Context Learning}

\begin{figure}[tb]
    \centering
    \includegraphics[width=1\linewidth]{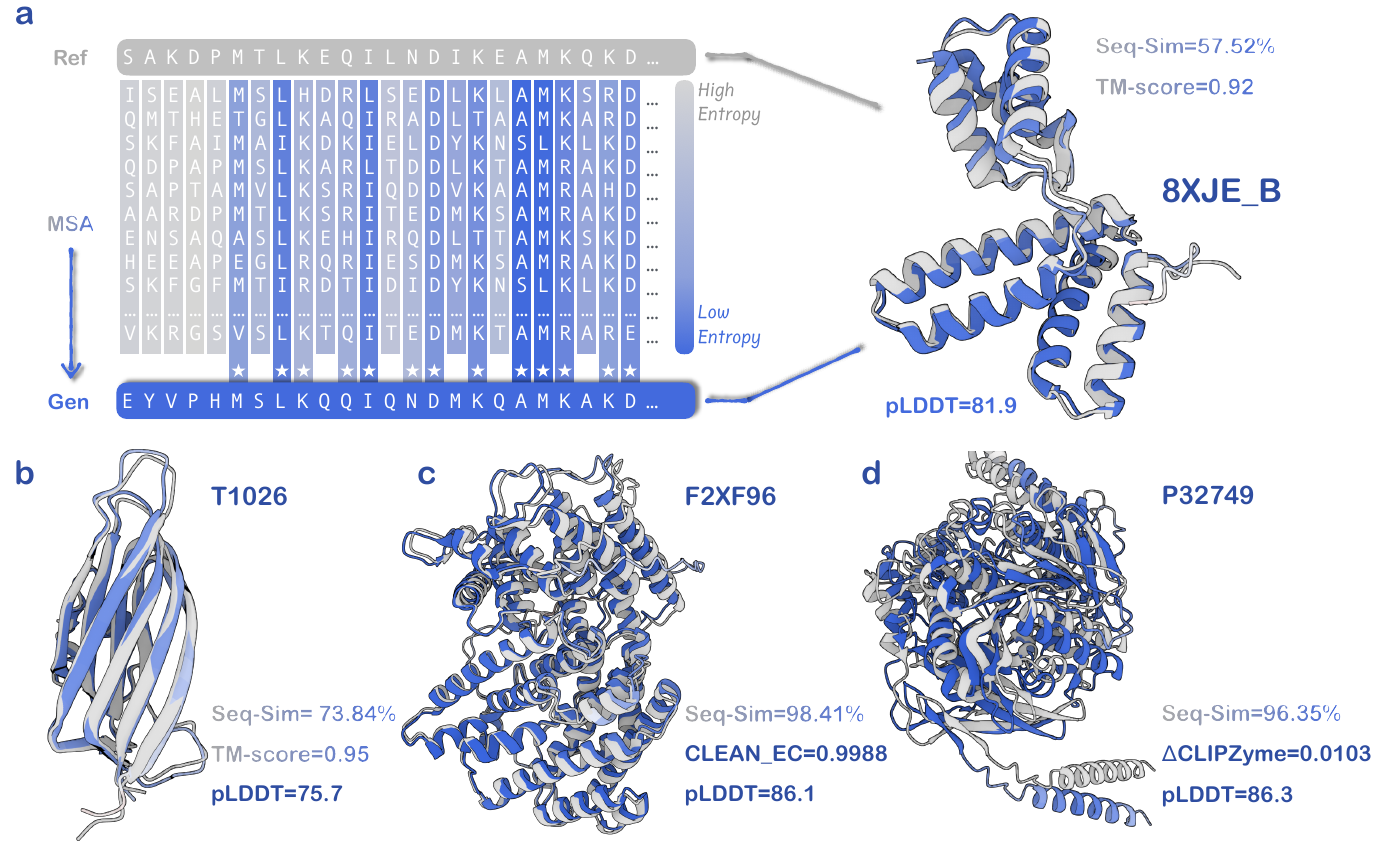}
    \caption{\textbf{Case studies validating in-context learning in \method.} \method preferentially conserves evolutionarily important positions—i.e., positions with low entropy (shown in \textbf{a} left, colored in \textcolor[HTML]{305FD2}{blue})—where the generated proteins more frequently match the wild-type sequence, compared to high-entropy (variable) positions (\textcolor[HTML]{B8B8B8}{gray}). The right side of \textbf{a}, as well as \textbf{b-d}, shows structural alignments and functional assessments between the generated and wild-type proteins. \textbf{(a, b)}: Structural similarity to the wild type, quantified by TM-score. \textbf{(c)}: Functional consistency based on EC Number, measured by CLEAN confidence. \textbf{(d)}: Catalytic activity preservation on a given reaction, measured by the change in CLIPZyme score ($\Delta$). In all panels, \textcolor[HTML]{305FD2}{blue} structures denote generated proteins, and \textcolor[HTML]{B8B8B8}{gray} structures denote the input wild type.}
    \label{fig:FewShotCase}
\end{figure}

We validate the in-context learning capability of \method through a set of in silico case studies. Each case examines whether the model can accurately extract and generalize structural or functional constraints from input exemplars—encoded as profiles—without access to explicit labels or structural supervision. We consider four representative case studies, each conditioned on a distinct profile, and generate novel proteins using \method. Structural properties of the output sequences are evaluated using ESMFold~\citep{lin2023evolutionary}, while functional predictions are made using CLEAN~\cite{yu2023enzyme} and CLIPZyme~\cite{mikhael2024clipzyme}. Seq-Sim of a generated protein $x'$ is computed as $\max_{x \in \mX} \frac{1}{m} \sum_{j=1}^{m} \mathbb{I}(x_j = x'_j)$, where $\mX$ is the input MSA and $m$ is the sequence length.

\paragraph{Structural Generalization from Random Homolog.}

The input consists of a randomly sampled protein family from CAMEO. As shown in \Cref{fig:FewShotCase}(a), \method generates novel proteins whose predicted 3D structures remain highly consistent with the input, as measured by structural alignment metric TM-score. The average predicted pLDDT exceeds 85, indicating confident and foldable outputs. Despite sequence-level novelty (low Seq-Sim to any input MSA sequence), the generated structures preserve the canonical fold, highlighting the model’s capacity to generalize conserved scaffolds.

\paragraph{Generalization from Orphan Protein.}

In \Cref{fig:FewShotCase}(b), we examine a challenging scenario where the input is an orphan protein—i.e., one with only a few known homologs and minimal representation in existing protein families. Such proteins are underrepresented in training data, making them difficult for models to generalize to. Despite this, \method successfully generates foldable variants that exhibit coherent tertiary structures with high pLDDT scores. This demonstrates that even sparse sequence profiles can serve as effective prompts, and that \method is capable of extrapolating structural patterns purely from weak but positionally localized evolutionary signals.

\paragraph{Function Retention—Enzyme with Original EC.}

We next assess functional generalization. \Cref{fig:FewShotCase}(c) presents results from prompting with an MSA containing enzymes annotated with EC 4.2.3.120 (beta-pinene synthase). The generated proteins are evaluated using CLEAN, a state-of-the-art model for EC number prediction. Over 80\% of the generations are assigned the correct EC number with high confidence, and the predicted functional entropy remains low, indicating strong specificity. The average predicted CLEAN confidence for generated sequences closely matches the input enzyme, demonstrating functional conservation through in-context prompting.

\paragraph{Catalytic Activity Conserved on Specific Reaction.}

Finally, we test whether \method can preserve catalytic activity for a chemically complex reaction. As shown in \Cref{fig:FewShotCase}(d), the input MSA consists of enzymes known to catalyze the transformation: \ce{[CH3][N+](CH3)(CH3)CH2CHO + H2O + O2 -> [CH3][N+](CH3)(CH3)CH2COOH + H2O2}. Using CLIPZyme, we quantify the catalytic plausibility of each generated sequence by computing the change in predicted activity scores relative to the input enzyme. We report a positive mean shift $\Delta$ in catalytic confidence across generations, demonstrating that the model does not merely preserve function—it can even improve functional alignment under certain scoring metrics.

Across these diverse scenarios, \method consistently demonstrates strong in-context learning capabilities. It synthesizes sequences that maintain both \textbf{structural fidelity} (as evidenced by high pLDDT and TM-score to prompt structure) and \textbf{functional specificity} (as validated by CLEAN and CLIPZyme predictions). Notably, this is achieved without any model fine-tuning or task-specific supervision, using only profile-based conditioning at inference time. These results underscore the utility of MSA profiles as compact, expressive prompts that unlock biologically meaningful generalization in protein generation.

\subsection{Wet Lab: \method Designs High-Activity Proteins} 

\begin{figure}[t]
    \centering
    \includegraphics[width=1.0\linewidth]{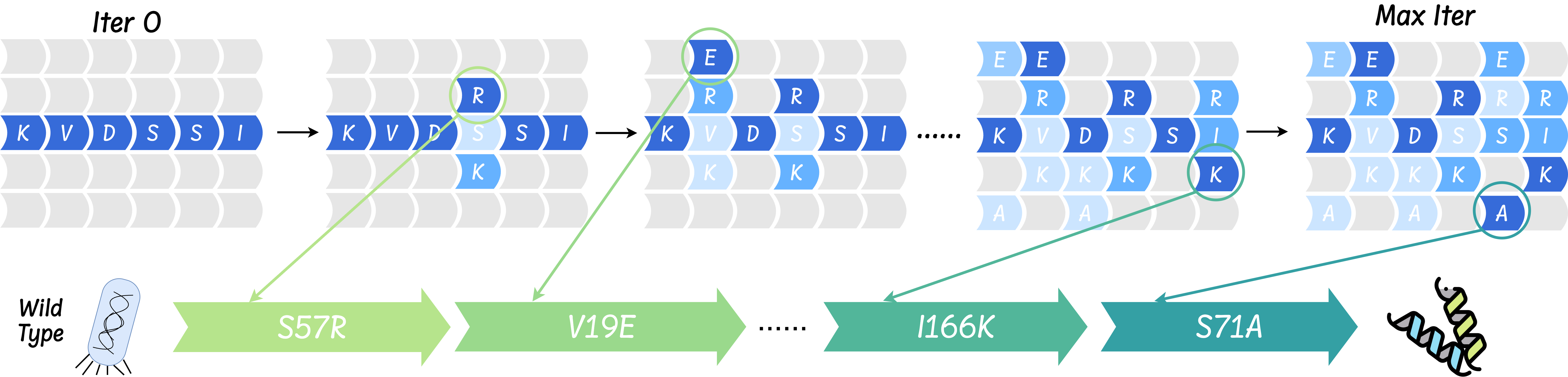}
    \caption{Generation Iterations and corresponding mutation process in wet-lab}
    \label{fig:wetlab_path}
\end{figure}
To validate the practical efficacy of \method's in-context learning, we experimentally engineered the AmeR transcriptional repressor~\citep{stanton2014genomic}. The results demonstrated that model-generated mutants enhanced the protein's activity by up to 50 fold compared to the wild type. This represents an approximate 77\% improvement over the best previously reported results.

\paragraph{Setup.} We targeted the AmeR protein, a TetR-family transcriptional repressor. In synthetic biology, AmeR is a widely used orthogonal molecular switch—a fundamental component for building predictable genetic circuits~\citep{stanton2014genomic}. High-performance switches are critical for sophisticated applications in metabolic engineering, biosensors and cell therapies. Our objective was to enhance AmeR's DNA binding affinity, thereby creating a more sensitive and reliable regulator.

\begin{wrapfigure}{r}{0.5\textwidth}
  \begin{center}
    \includegraphics[width=0.48\textwidth]{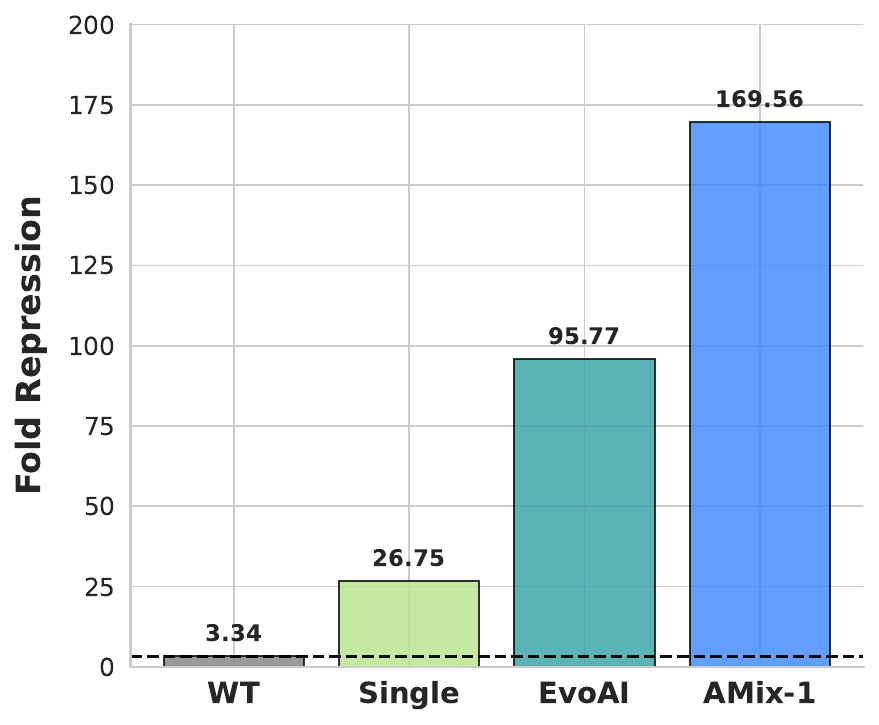}
  \end{center}
  \caption{Best Fold Repression Score achieved through different Methods. WT for original wildtype sequence, single stands for a mutation with only one residue different.}
  \label{fig:wetlab_result}
\end{wrapfigure}

Using our \method-650M model, we generated 40 candidate sequences, each with ten or fewer mutations. The fitness of these variants was quantified using a fluorescent reporter assay, where higher DNA binding affinity results in stronger repression of a fluorescent protein, measured as ``fold repression''. A higher fold repression value indicates superior function. The performance of \method was benchmarked against three baselines: (1) Wild Type (WT); (2) a Single-Mutant Library; and (3) EvoAI~\citep{ma2025evoai}, a state-of-the-art directed evolution method. Illustration of wet lab verification process guided iteratively by \method is shown in \Cref{fig:wetlab_path} while detailed protocols are provided in \Cref{app:wet-lab}.

\paragraph{Results.} The full experimental results are summarized in \Cref{fig:wetlab_result}. The superior performance of the \method-designed variants highlights our model's key strengths. It demonstrates a powerful ability to navigate the vast sequence-function landscape, identifying not just individual high impact mutations but also beneficial combinatorial effects. This capability allows \method to discover highly active sequences inaccessible to methods that rely on local exploration or the recombination of known variants.

    \section{Test-Time Scaling}
\label{sec:tts}

Test-time scaling (TTS) enhances model performance by allocating additional computational resources during inference time~\cite{snell2024scaling,openai2024openaio1card,muennighoff2025s1}.
In this section, we introduce \ttsmethodt, an evolutionary test-time scaling algorithm that leverages \method as a proposer and an external verifier to iteratively generate and select candidate protein sequences under evolutionary constraints. 
By integrating either task-specific \textit{in silico} reward functions or experimental feedback from assays, \ttsmethod enables efficient directed protein evolution without model fine-tuning.
Extensive experiments show that \ttsmethod consistently outperforms \method's in-context learning and strong baseline methods across all six design tasks, highlighting its potential for real-world applications in protein engineering.

\begin{DefinitionBox}
\textbf{Key Result} \ding{175}: The protein design ability of \method is scalable in test time with increasing verification budgets, and is universal to apply to a variety of directed evolution tasks with plug-in verification.
\end{DefinitionBox}

\subsection{Scaling \method with Evolutionary Algorithm}

\begin{figure}[tb]
    \centering
    \includegraphics[width=1\linewidth]{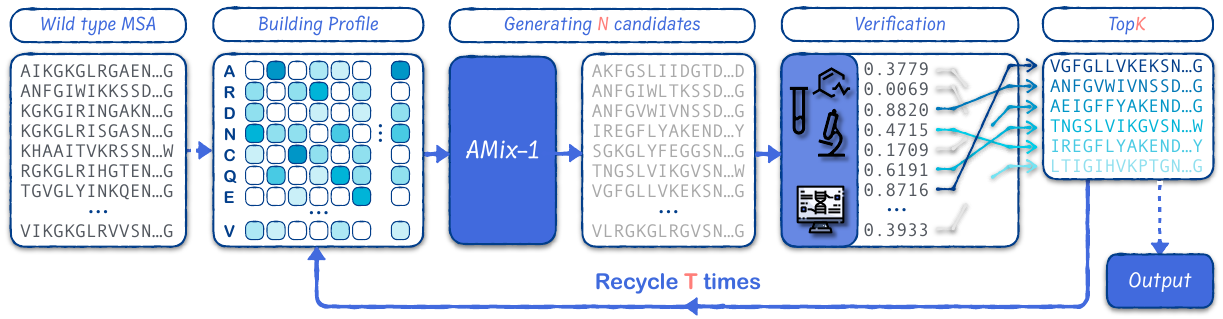}
    \caption{Schematic of the \method test-time scaling loop. Starting from an initial MSA of the wild-type sequence, a position-wise frequency profile $\mP^{(0)}$ is computed and serves as input to the \textbf{proposer} (\method), which generates a batch of $N$ candidate sequences. An external \textbf{verifier} $R(\cdot)$ scores each candidate and selects the top-$k$ variants, from which a refined profile $\mP^{(t+1)}$ is constructed. This propose-verify-update cycle repeats for $T$ rounds, yielding progressively optimized protein designs without any parameter updates.
}
    \label{fig:tts_workflow}
\end{figure}

\ttsmethod is structured as a proposer-verifier framework, where \method serves as the proposer that generates candidate protein variants, and a verifier evaluates their fitness to guide the next round of selection.
We here make no assumptions about whether the verifier is a computational metric or a wet-lab assay, ensuring broad compatibility and enabling a (simulated) lab-in-the-loop algorithm.
Notably, unlike typical test-time scaling methods in large language models that primarily focus on increasing computational budgets, we shift the focus toward verification budget~\cite{ma2025inference}.
In real-world protein engineering, experimental verification via wet-lab assays is significantly more costly and time-consuming than computational inference, often taking weeks or months to complete. 
This verification step is widely regarded as a major bottleneck.
Therefore, in our test-time scaling study, we aim to devise an evolutionary algorithm that consistently improves \method's performance by investing increasing verification budget.

Operationally, we instantiate \ttsmethod as an evolutionary algorithm guided by a verifier that estimates fitness, inspired by the principles of natural selection (see \Cref{fig:tts_workflow}).
Given a wild-type MSA, we encode it as a profile, a categorical distribution over amino acids, and use \method to generate a diverse batch of candidate variants.
Leveraging its in-context learning capability in \Cref{sec:few-shot}, \method produces variants that inherently preserve the structural and functional characteristics of the input.
Each variant is then evaluated by a verifier that assigns a fitness score. Owing to \method’s stochastic nature, the generated sequences span a range of fitness levels, with some surpassing all previous variants.
High-fitness candidates are retained, while low-fitness ones are discarded. The surviving variants are used to construct a pseudo-MSA, which is then compressed into a new profile.
This profile serves as the input to the next round of variant generation.
By iteratively applying this propose-verify-update cycle, \ttsmethod progressively refines its outputs, producing increasingly fit protein sequences in a test-time scalable fashion.
A detailed algorithm is provided in \Cref{alg:tts}.

An underlying mechanism of \ttsmethod closely parallels reinforcement learning.
On one hand, the inherent stochasticity of the probabilistic model encourages exploration, allowing the generation process to sample variants beyond the input proteins.
On the other hand, \method conditions on previously selected high-fitness variants, using them as prompts to implicitly regularize generation, which exploits shared successful patterns among these variants as prior knowledge to guide future rounds.
This duality enables \ttsmethod to balance exploration and exploitation so that it explores novel sequence space while exploiting learned evolutionary patterns.
As a result, \ttsmethod delivers robust and test-time scalable performance in protein design.


\subsection{A Proposal Distribution Perspective}
\begin{figure}
    \centering
    \includegraphics[width=0.9\linewidth]{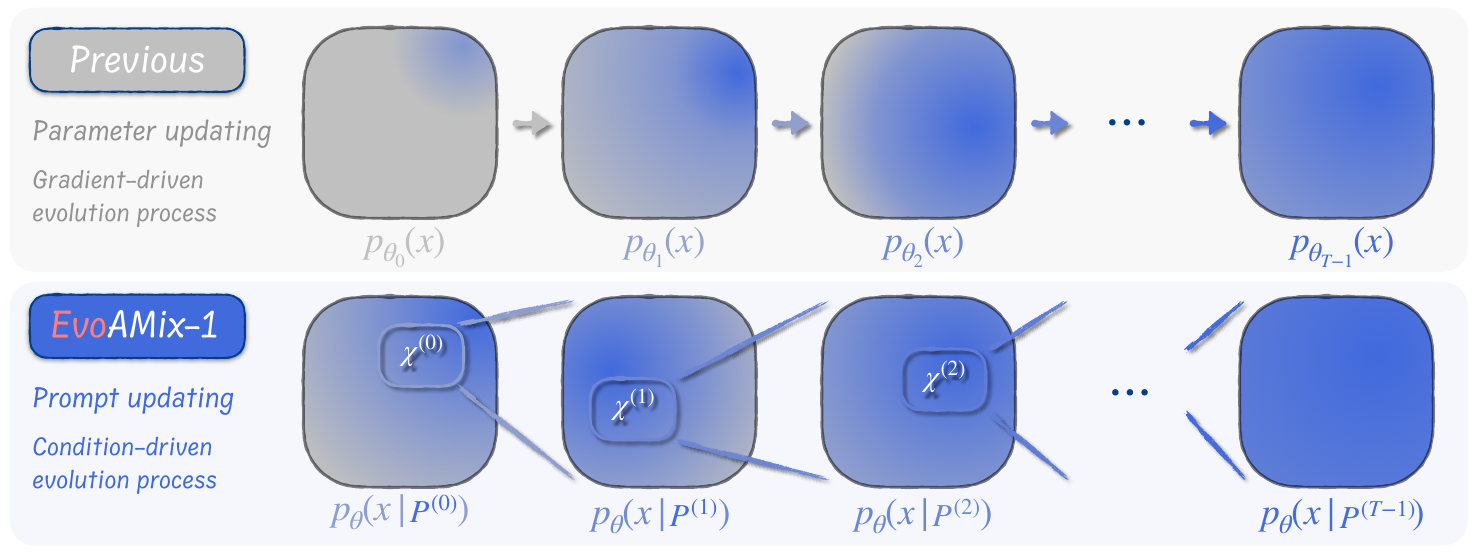}
    \caption{Proposal distribution shifting during the test-time scaling process.
(1) Top: Previous methods typically adopt a gradient-driven evolution process, in which the model parameters $\theta$ are iteratively updated. During optimization, the proposal distribution shifts as the model becomes increasingly tailored to generate samples that align with feedback from the verifier in the previous round. This approach modifies the underlying model to improve precision on newly verified data points.
(2) Bottom: In contrast, \ttsmethod utilizes a condition-driven evolution strategy through prompt updating rather than parameter tuning. At each iteration, \ttsmethod identifies the most promising sequences and updates the in-context examples (e.g., MSAs or profiles) that serve as prompts to the foundation model. This iterative prompt refinement effectively reshapes the proposal distribution without altering the model itself, progressively narrowing the generation focus toward regions of sequence space favored by the verifier.}
    \label{fig:tts_theory}
\end{figure}
We now turn to a more formal discussion of the mechanism underlying \ttsmethod, framing it through the lens of \emph{proposal distributions} as shown in \Cref{fig:tts_theory}. 
This perspective not only clarifies the optimization dynamics of our method, but also unifies traditional evolutionary algorithms in computational biology under a common framework, highlighting key distinctions between those approaches and \ttsmethod.

As proposed in \citet{snell2024scaling}, test-time scaling (TTS) algorithms can be interpreted through a proposer-verifier framework, where the proposer samples candidate solutions from a distribution, and the verifier evaluates them, selecting those with the highest fitness. Formally, this can be modeled as a sequence of proposal distributions, denoted as $p^{(0)}(\mathbf{x}), p^{(1)}(\mathbf{x}), \ldots, p^{(T-1)}(\mathbf{x})$, where $p^{(t)}(\mathbf{x})$ represents the proposal distribution at the $t$-th optimization round, and $T$ is the maximum number of iterations. 
At each step, the algorithm samples candidate proteins from $p^{(t)}(\mathbf{x})$ and updates the proposal distribution to $p^{(t+1)}(\mathbf{x})$ by a combined signal of the last distribution, the generated candidates, and feedback from the verifier. 
This iterative process reflects the dynamic adaptation of the proposal distribution in response to fitness signals.

Many conventional methods, such as ALDE~\cite{yang2025active} and EVOLVEpro~\cite{jiang2024rapid}, can also be interpreted within this framework. 
These methods adopt a parametric predictor to shape a proposal distribution and generate new protein variants with importance sampling.
The sampled protein are evaluated via \textit{in silico} or wet-lab assays and the evaluation results are then used to retrain the predictor, producing a new proposal distribution for the next round. 
This yields a sequence of parameterized distributions $p_{\theta^{(0)}}(\mathbf{x}), p_{\theta^{(1)}}(\mathbf{x}), \ldots, p_{\theta^{(T-1)}}(\mathbf{x})$, where $\theta^{(t)}$ represents the model parameters at iteration $t$. The optimization process is thus driven by continual updates to the model parameters to better fit the observed feedback.

In contrast, \ttsmethod adopts a fundamentally different philosophy: rather than updating model parameters, it constructs proposal distributions through in-context exemplars. At each round, \method takes a set of MSAs or their profile as a prompt, treating it as the input condition to the protein foundation model. The foundation model then samples neighboring sequences, effectively defining a new conditional proposal distribution. This yields a sequence of context-dependent distributions: $p(\mathbf{x}|\mP^{(0)}),p(\mathbf{x}|\mP^{(1)}),\ldots,p(\mathbf{x}|\mP^{(T-1)})$, where $\mP^{(t)}$ is the prompt at round $t$, and $p(\mathbf{x})$ is the frozen base distribution defined by the pretrained protein foundation model. This approach assumes that the model has already learned a rich, generalizable manifold of natural proteins and contains comprehensive structural and functional knowledge.

Rather than fine-tuning model weights, \ttsmethod embraces a prompting-based methodology, akin to in-context learning~\cite{brown2020language, dong2022survey} and chain-of-thought prompting in large language models~\cite{wei2022chain}. 
This allows for adaptive behavior during optimization without the need for costly retraining, aligning with the broader trend in foundation models of extracting knowledge through flexible conditioning rather than parameter updates.

\subsection{\method Exhibits Robust Test-Time Scaling across Diverse Tasks}
\begin{figure}[tb]
    \centering
    \includegraphics[width=1\linewidth]{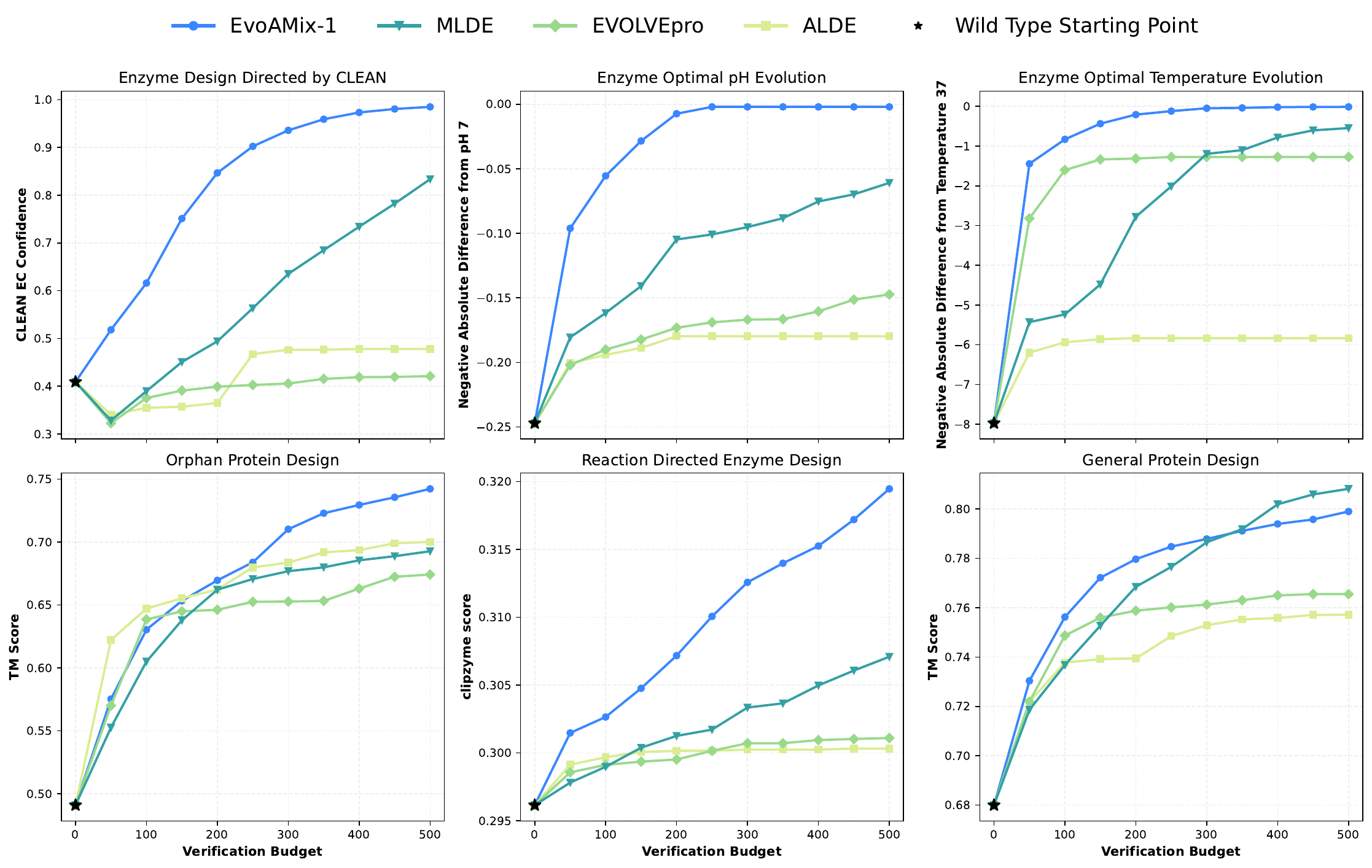}
    \caption{Test-time scaling performance of \method across six in silico directed-evolution benchmarks. Each curve shows mean top-5 task score as a function of cumulative verifier calls (proxy for experimental throughput) for structural, biophysical, and functional design objectives. \method exhibits monotonic performance gains with increasing inference compute, outperforms or matches state-of-the-art baselines in five out of six tasks, and avoids the premature plateaus observed in methods with restricted mutation schemes.}
    \label{fig:tts_result}
\end{figure}

We applied \ttsmethod to a series of protein design tasks.
The results in \Cref{fig:tts_result} show that it consistently exhibits strong test-time scaling behavior across all tasks—namely, as the number of verifier calls increases, the model continues to produce progressively higher-quality designs.
Specifically, we evaluate \ttsmethod across three representative classes of protein design tasks:
\begin{itemize}
    \item \textbf{Identifying structurally consistent family members}. This task requires the model to generate protein sequences whose structures are more similar to a given reference. To evaluate this, we use ESMFold to predict the structures of generated sequences and compute their TM-scores against the reference structure. The TM-score serves as the output of the verifier in this task. This task covers both orphan proteins (with limited homologous information) and general proteins (with dense homology), allowing us to assess the robustness of \ttsmethod under varying levels of evolutionary context.
    \item \textbf{Optimization of biophysical properties}. This task requires the model to iteratively generate variants starting from a wild-type sequence, guiding the design toward desired target properties. In this work, we focus on two representative biophysical attributes of enzymes: optimal temperature and pH. We use Seq2Topt~\cite{qiu2025seq2topt} and Seq2pHopt~\cite{qiu_seq2topt_2023} as task-specific verifier models to evaluate these properties.
    \item \textbf{Functional reprogramming}. This task aims to modify or enhance the function of a wild-type protein. Specifically, we focus on reprogramming enzymatic functions by targeting either EC number annotations or catalytic activity toward specific chemical reactions. We use CLEAN~\cite{yu2023enzyme} and CLIPZyme~\cite{mikhael2024clipzyme} as functional verifiers for these two respective objectives.
\end{itemize}
In addition, we selected a set of strong directed evolution algorithms as baselines for comparison, including:
\begin{itemize}
    \item \textbf{ALDE}~\cite{yang2025active} is an iterative protein engineering strategy that combines machine learning with Bayesian optimization to efficiently explore the epistatic sequence space and identify high-fitness protein variants. However, it requires manual specification of mutation sites, which significantly constrains its exploration space.
    \item \textbf{EVOLVEpro}~\cite{jiang2024rapid} is an active learning framework that integrates a pretrained protein language model with a regression model to guide in silico directed evolution and discover high-activity protein variants. Notably, EVOLVEpro limits its search space to single-point mutations, reducing its capacity to explore more diverse or combinatorial sequence variations.
    \item \textbf{MLDE}~\cite{tran2024protein} utilizes ESM language models and combines importance-based masking with random masking strategies to progressively introduce mutations into the wild-type sequence. It automatically generates and optimizes protein variants aimed at enhancing target functionality.
\end{itemize}

We report performance with respect to the verification budget, i.e., the number of verifier calls, as this directly reflects wet-lab throughput in real-world protein design scenarios.
This perspective offers practical insight for lab-in-the-loop frameworks.
More implementation details of these experiments can be found in \Cref{app:tts}. We now proceed to a more detailed discussion of the experimental results.

\paragraph{\ttsmethod is a universal protein optimizer.}
Across all directed evolution tasks, \ttsmethod demonstrates strong potential as a general-purpose optimizer for protein design under arbitrary objective functions. 
In contrast to our baseline approaches that rely on specialized training with task-specific environmental feedback, \ttsmethod operates without requiring retraining for each protein design dataset. Instead, it treats environmental feedback as a modular component of the optimization loop. 
A particularly appealing feature of \ttsmethod is that the feedback mechanism, whether derived from \textit{in silico} metrics or experimental assays, can be seamlessly integrated as a plug-in. 
The success across diverse tasks highlights \ttsmethod’s flexibility and effectiveness, establishing it as a universal protein optimizer.

\paragraph{\ttsmethod consistently produces strong final designs.}
In 5 out of 6 tasks, \ttsmethod achieved the best design performance. 
It was slightly outperformed by MLDE only in the general protein design task, yet it still significantly outperformed all other baseline methods.
This highlights the generality of \method, demonstrating its ability to quickly adapt to diverse types of protein design tasks.

\paragraph{\ttsmethod exhibits fast scalability.}
Across all evaluated tasks, \ttsmethod demonstrates rapid optimization performance, achieving the fastest improvement rate in 5 out of 6 benchmarks. 
This scalability advantage stems from its flexible design: unlike traditional approaches that constrain each evolutionary round to a predefined set of mutation sites, 
\ttsmethod enables unrestricted exploration over the entire sequence space. 
This broader search strategy facilitates more efficient discovery of high-performing protein variants, accelerating the overall optimization process.

\paragraph{\ttsmethod escapes local optima.}
In all tasks, \ttsmethod continued to improve without observable performance plateaus, as long as the metric had not yet reached its theoretical upper bound.
In comparison, ALDE and EVOLVEpro showed noticeable performance plateaus in several tasks, indicating limited capacity for continued improvement.
This limitation is primarily due to the manual constraints imposed by both methods on mutation types and allowable mutation sites.
Unlike these methods, \ttsmethod does not rely on manually defined constraints, enabling broader and more flexible exploration of the sequence space.

\paragraph{Ablation studies confirm key design choices.}
To better understand the contribution of each component in \ttsmethod, we conduct targeted ablation experiments presented in \Cref{tts_ablation}. 
These studies reveal that removing either the pretrained generative model or the iterative profile refinement mechanism leads to substantial drops in both reward optimization and biological plausibility. 
Notably, substituting \method with random mutation causes the model to quickly drift away from structurally and functionally coherent solutions, while disabling prompt updates prevents convergence to high-fitness regions. 
These findings validate the importance of valid generation and adaptive conditioning in realizing effective test-time scaling.

    \section{Conclusion}

In this work, we propose a pathway methodology for training a scalable and emergent protein language model based on the Bayesian Flow Network (BFN) for scalable protein design. 
This methodology addresses key challenges in developing protein foundation models, inspired by advances in large language models, namely, scaling laws, emergent capabilities, in-context learning, and test-time scaling. 
Demonstrating scalability across data, model size, and computational resources, we train a 1.7 billion \method, in which sequence and structural understanding progressively emerge during training. 
Leveraging multiple sequence alignments, we introduce an in-context learning mechanism that enables \method to generate novel proteins conditioned on a small number of examples, while preserving structural similarity to the input proteins. 
Wet-lab experiments further validate that \method significantly improves the design of high-activity AmeR variants. 
In addition, we present an evolutionary test-time scaling algorithm that enhances \methodt’s applicability to protein-directed evolution, and validate its performance using a variety of \textit{in silico} target metrics, simulating a lab-in-the-loop environment to facilitate future real-world integration.

\paragraph{Limitation and Future Work} 
Although \method demonstrates promising results, we recognize several limitations in its current form.
As the field moves toward protein foundation models that jointly learn from both sequence and structure, \method remains a purely sequence-based learning paradigm, leaving a large amount of structural data unutilized, despite growing evidence that structural information significantly enhances protein understanding. 
Additionally, the test-time scaling algorithm for \method has so far only been validated on simulated tasks using \textit{in silico} metrics, lacking experimental assays for real-world verification.
As part of future work, we are actively extending \method by integrating structural components and applying it to a broader range of real-world experimental assays.

    \newpage
\section*{Contributions}

\textbf{Project Lead}

Jiangtao Feng$^{1,2,3}$

\textbf{Scaling Law \& Emergent Ability}

Changze Lv$^{1,2,5}$, Lihao Wang$^{1,2}$, Dongyu Xue$^{1,2}$

\textbf{In-Context Learning \& Test-time Scaling}

Jiang Zhou$^{6,7,1,2}$, Siyu Long$^{2,3}$, Dongyu Xue$^{1,2}$

\textbf{Evaluation}

Zherui Zhang$^{1,8}$, Yuchen Cai$^{1,5}$, Zhiqiang Gao$^{1}$

\textbf{Wet Experiment}

Yu Pei$^{2,3}$, Hao Wang$^{2,3}$, Ziyuan Ma$^{4}$, Shuyi Zhang$^{4}$

\textbf{Infrastructure}

Lihao Wang$^{1,2}$, Jiakai Hu$^{1}$

\textbf{Other Contributers}

Chaochen Gao$^{9}$, Jingjing Gong$^{2,3}$, Yuxuan Song$^{2,3}$, Xiaoqing Zheng$^{5}$, Deyi Xiong$^{6}$, Lei Bai$^{1}$, Wanli Ouyang$^{1}$, Ya-Qin Zhang$^{3}$, Wei-Ying Ma$^{3,10}$, Bowen Zhou$^{1,4}$

\textbf{Correspondence}

Hao Zhou$^{1,2,3}$

\subsection*{Affiliation}

$^1$Shanghai Artificial Intelligence Laboratory

$^2$Generative Symbolic Intelligence Lab (GenSI), Tsinghua University

$^3$Institute for AI Industry Research (AIR), Tsinghua University

$^4$Tsinghua University

$^5$Fudan University

$^6$Tianjin University

$^7$Georgia Institute of Technology

$^8$Beijing University of Posts and Telecommunications

$^9$University of Chinese Academy of Sciences

$^{10}$City University of Hong Kong

\section*{Acknowledgments}

We thank DeepLink Team for their infrastructure support. 
This work is supported by Shanghai Artificial Intelligence Laboratory. 
    
    \clearpage
    \bibliographystyle{unsrtnat}
    \bibliography{refs}

    \clearpage
    \beginappendix
    
\section{Pretrainig Algorithm}\label{app:pretrainalgo}
\subsection{Training}
The continuous-time discrete Bayesian flow loss, following the ProfileBFN framework~\cite {jingjing2025ProfileBFN}, is adopted for model training.
\begin{align}\label{equ:contdiscreteloss}
\mathcal{L}(\vphi) &= \mathbb{E}_{\mathbf{x} \sim p_{\text{data}}}\mathbb{E}_{\Pi_{i=1}^n p_\text{s}(\mathbf{y}_i|\mathbf{x},\alpha_i)}  D_{KL} \left(p_\text{s}(\mathbf{y}_i|\mathbf{x},\alpha_i)||p_\text{r}(\mathbf{y}_i\mid \vtheta_{i-1},\vphi,\alpha_i)\right) \\
&= \frac{1}{2}\beta'(t)K\lVert\vphi(\vtheta_{i-1}) - \mathbf{e}(\mathbf{x})\rVert^2 \\
&= \beta_1tK\lVert\vphi(\vtheta_{i-1}) - \mathbf{e}(\mathbf{x})\rVert^2
\end{align}

\begin{algorithm}
\caption{Discrete Bayesian Flow Loss}
\label{alg:discretebfnloss}
\begin{algorithmic}[1]
\Require data points $\mathbf{x}$, output distribution neural network $\vphi$, 
\Ensure $\mathcal{L}(\vphi)$
\State $t \sim U(0, 1)$
\State $\mathbf{y}_{t} \sim \mathcal{N}\bigl(\beta_1t^2 \left(K \mathbf{e}(\mathbf{x}) - \mathbf{1}\right),  \beta_1t^2K \mathbf{I}\bigr)$
\State $\boldsymbol{\theta_t} \gets \texttt{softmax}(\mathbf{y}_t)$
\State $\mathcal{L}(\vphi) = \beta_1tK \left\lVert \phi\left(\boldsymbol{\theta}_t\right) - \mathbf{e}(\mathbf{x})  \right\rVert^2$
\State \Return $\mathcal{L}(\vphi)$
\end{algorithmic}
\end{algorithm}

\subsection{Inference}

\begin{algorithm}
\caption{Noise Reduced Sampling}
\label{alg:noise_reduced_sampling}
\begin{algorithmic}[1]
\Require $N$ steps, output distribution neural network $\vphi$
\Ensure inference sample $\mathbf{x}$
\State $\mathbf{z} \sim \mathcal{N}(\mathbf{0}, \mathbf{I})$
\State $\mathbf{y}^{(0)} \gets \mathbf{0}$
\For{$i = 0 \ldots N-1$}
    \State $t \gets \frac{i}{N}$ $j \gets \frac{i + 1}{N}$
    \State $\boldsymbol{\theta}_{t} \gets \texttt{softmax}(\mathbf{y}_{t})$
    \State $\beta(t) \gets \beta_1 t^2$
    \State $\mathbf{y}_{j} \gets \beta(j) \left(K \phi\left(\boldsymbol{\theta}_{t}\right) - \mathbf{1}\right) + \sqrt{K \beta(j)} \mathbf{z}$
\EndFor
\State $\boldsymbol{\theta}_{1} \gets \texttt{softmax}(\mathbf{y}_{1})$
\State $\mathbf{x} = \texttt{argmax}(\phi\left(\boldsymbol{\theta}^{(1)}\right))$ 
\State \Return $\mathbf{x}$
\end{algorithmic}
\end{algorithm}
\newpage
\section{Performance of Prediction Tasks}
\begin{table}[htbp]
\centering
\caption{\method demonstrates robust capabilities across diverse protein prediction tasks, indicating a strong comprehension of protein. To ensure a fair comparison, we evaluated various models with 650 million parameters, all initialized from ESM2 weights. *: protein structure is provided. $\dagger$: results are quoted from SaProt~\cite{su2023saprot}.  The best scores are highlighted in \textbf{bold} and the second-best scores are \underline{underlined}. }
\label{tab:protein_prediction}
\resizebox{\linewidth}{!}{
\begin{tabular}{lccccccccc}
\toprule
\multirow{3}{*}{Model(650M)} & \multirow{2}{*}{Thermostability} & \multirow{2}{*}{HumanPPI} & \multirow{2}{*}{Metal Ion Binding} & \multirow{2}{*}{EC} & \multicolumn{3}{c}{GO} & \multicolumn{2}{c}{DeepLoc} \\
\cmidrule(lr){6-8} \cmidrule(lr){9-10}
 &  &  &  &  & MF & BP & CC & Subcellular & Binary \\
\cmidrule(lr){2-2} \cmidrule(lr){3-3} \cmidrule(lr){4-4} \cmidrule(lr){5-5} \cmidrule(lr){6-8} \cmidrule(lr){9-10}
 & Spearman's $\rho$ & ACC(\%) & ACC(\%) & Fmax & Fmax & Fmax & Fmax & ACC(\%) & ACC(\%) \\
\midrule
ESM-2 $\dagger$ & 0.680 & 76.67 & 71.56 & 0.877 & 0.668 & 0.345 & 0.411 & 82.09 & 91.96 \\
SaProt* & \textbf{0.724} & 86.41 & \textbf{75.75} & \textbf{0.884} & 0.678 & 0.356 & 0.414 & \textbf{85.57} & \underline{93.55} \\
DPLM  & 0.695 & 86.41 & \underline{75.15} & 0.875 & \underline{0.680} & 0.357 & 0.409 & 84.56 & 93.09 \\
DPLM-2  & 0.714 & 84.44 & 74.28 & \underline{0.878} & \underline{0.680} & \underline{0.359} & 0.411 & 82.98 & \textbf{93.64} \\
\midrule
 \method & \underline{0.718} & \textbf{89.13} & 71.13 & 0.875 & \textbf{0.682} & \textbf{0.495} & \textbf{0.482} & \underline{84.62} & 93.31 \\
\bottomrule
\end{tabular}
}
\end{table}
\section{Details of CAMEO-20}\label{app:cameo20}
In this section, we will show detailed information about CAMEO-20, which we applied to evaluate the performance of \method models on downstream tasks.

\begin{table}[htp]
\footnotesize
\caption{Detailed information of each protein for the CAMEO-20 dataset, partially from ProfileBFN \cite{jingjing2025ProfileBFN}}
\centering
\label{tab:cameo_details}
\resizebox{\textwidth}{!}{
\begin{tabular}{ccccc}
\toprule
PDB & Chain & Length & Title \\
\midrule
8F9R & A & 501 & Rabbit sialic acid esterase \\
8JDH & A & 166 & Crystal structure of anti-CRISPR AcrIF25 \\
8JGO & A & 535 & Crystal structure of Deinococcus radiodurans exopolyphosphatase \\
8JRB & A & 597 & Structure of DNA polymerase 1 from Aquifex pyrophilus \\
8QL0 & A & 693 & Structure of human PAD6 Phosphomimic mutant V10E/S446E, apo  \\
8QVC & B & 100 &  Deinococcus aerius TR0125 C-glucosyl deglycosidase (CGD), wild type crystal cryoprotected with glycerol \\
8S4S & A & 145 & PrgE from plasmid pCF10  \\
8SUF & A & 1007 &  The complex of TOL-1 ectodomain bound to LAT-1 Lectin domain \\
8SUF & E & 114 &  The complex of TOL-1 ectodomain bound to LAT-1 Lectin domain \\
8SW5 & C & 47 & Protein Phosphatase 1 in complex with PP1-specific Phosphatase targeting peptide (PhosTAP) version 1  \\
8UAI & B & 494 &  Crystal structure of hetero hexameric hazelnut allergen Cor a 9 \\
8V8P & A & 231 &  Sorghum Chalcone Isomerase \\
8WEX & A & 468 &  Crystal structure of N-acetyl sugar amidotransferase from Legionella pneumophila \\
8WG0 & D & 100 &  Crystal structure of GH97 glucodextranase from Flavobacterium johnsoniae in complex with glucose \\
8WOP & A & 100 &  Crystal structure of Arabidopsis thaliana UDP-glucose 4-epimerase 2 (AtUGE2) complexed with UDP, wild-type \\
8WTB & B & 187 &  Crystal structure of McsA/McsB complex truncated by chymotrypsin \\
8XJE & B & 153 &  Crystal structure of the YqeY protein from Campylobacter jejuni \\
8XJG & A & 153 & Crystal structure of the YqeY protein from Vibrio parahaemolyticus \\
9B1R & A & 562 &  Functional implication of the homotrimeric multidomain vacuolar sorting receptor 1 from Arabidopsis thaliana \\
9BCZ & A & 644 &  Chicken 1-phosphatidylinositol 4,5-bisphosphate phosphodiesterase zeta-1 (PLCZ1) in complex with calcium and phosphorylated threonine \\
9F63 & A & 572 &  Crystal structure of Saccharomyces cerevisiae pH nine-sensitive protein 1 (PNS1) \\
\bottomrule
\end{tabular}  
}
\end{table}

\section{Further Discussion on Scaling Laws}
\subsection{Unscalability when Extreme Noise Levels}\label{app:unscaling}

\begin{figure}[htp]
\centering
\includegraphics[width=0.98\linewidth]{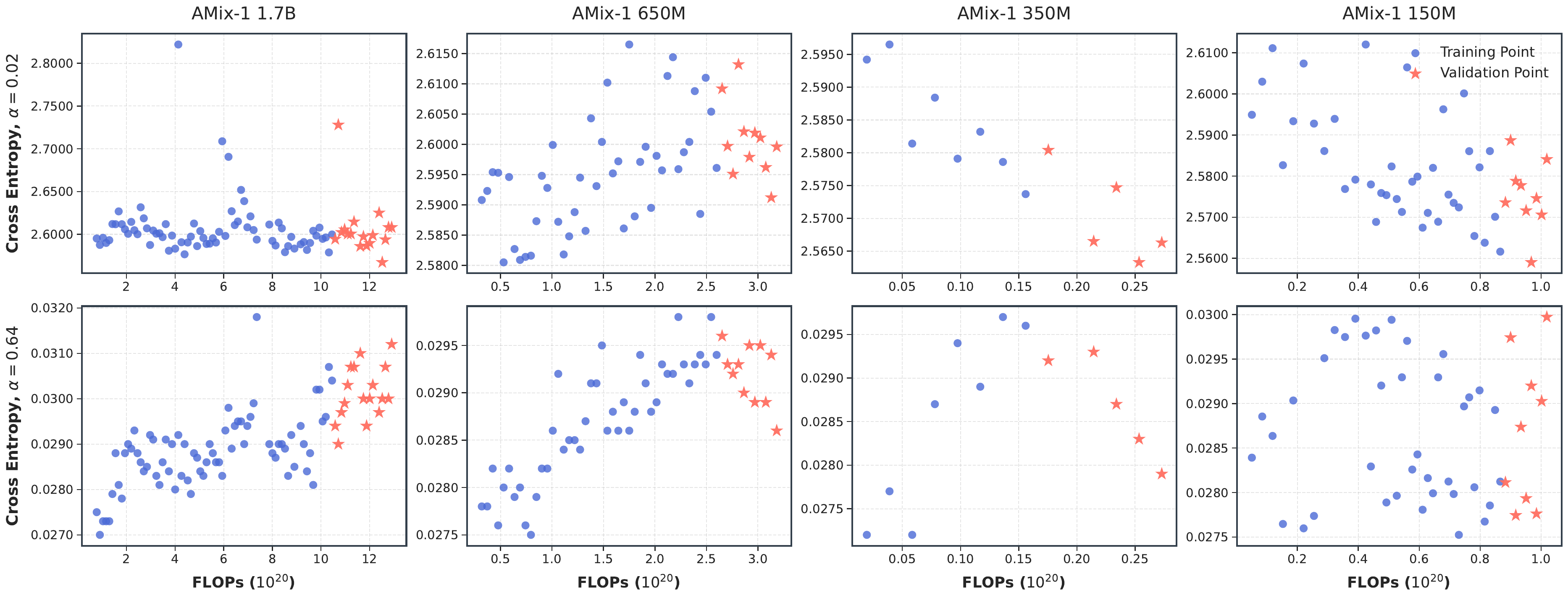}
\caption{Cross entropy and FLOPs may be unpredictable under several noise levels.}
\label{fig:unscaling}
\end{figure}

We show the variation of cross-entropy $\mathcal{L}_{CE}$ under noise level $\alpha=0.02$ and $0.64$ in \Cref{fig:unscaling}, and we find that there is no scaling behavior at all on these settings.

\subsection{Mean Relative Error} \label{app:mre}

We utilize the mean relative error (MRE) to assess the goodness of scaling laws, defined as:
\begin{equation}
\text{Fitting MRE} = \frac{1}{|\mathcal{D}_\text{fit}|} \sum_{(x, y) \in \mathcal{D}_\text{fit}} \left| \frac{f_\theta(x) - y}{y} \right|, \quad 
\text{Valid MRE} = \frac{1}{|\mathcal{D}_\text{val}|} \sum_{(x, y) \in \mathcal{D}_\text{val}} \left| \frac{f_\theta(x) - y}{y} \right|.
\end{equation}
Here, $f_\theta(x)$ is the predicted loss given model size $N$ and training tokens $D$, and $y$ is the actual cross-entropy.

\subsection{Parametric Scaling Laws under Several Noise Levels}\label{app:param_scale}

In addition to the noise levels reported in \Cref{tab:paramscaling}, we evaluate a broader range of noise settings, as summarized in \Cref{tab:paramscaling_left}.
These additional results further reinforce our conclusion that \method models exhibit consistent scalability when trained under appropriately chosen noise levels.


\begin{table}[htb]
\centering
\caption{Parametric scaling laws under noise level $\alpha\in \{0.06,0.08, 0.48, \text{ and }0.64\}$. Note that both $N$ and $D$ are measured in millions.}
\resizebox{0.8\linewidth}{!}{
\begin{tabular}{c| c c c c c | c c | c}
\toprule
\multicolumn{1}{c}{\textbf{Noise Level}} & \multicolumn{5}{c|}{\textbf{Parameters}} & \multicolumn{2}{c|}{\textbf{MRE (\%)}} & \multirow{2}{*}{\textbf{Scalable}} \\
\cmidrule(lr){1-1} \cmidrule(lr){2-6} \cmidrule(lr){7-8}
$\alpha$ & $E$ & $A$ & $B$ & $n$ & $d$ & Fitting & Valid & \\
\midrule
$0.06$ & $1.4764$ & $0.5914$ & $5.4300$ & $0.5687$ & $0.2865$ & $0.55$ & $1.84$ & \textcolor[HTML]{4A69D6}{\cmark}\\ 
\midrule
$0.08$ & $1.1630$ & $0.0187$ & $3.1436$ & $0.0163$ & $0.2093$ & $0.66$ & $1.28$ & \textcolor[HTML]{4A69D6}{\cmark}\\ 
\midrule
$0.48$ & $0.0583$ & $0.0231$ & $0.3050$ & $0.2006$ & $0.2563$ & $1.10$ & $1.82$ & \textcolor[HTML]{4A69D6}{\cmark}\\ 
\midrule
$0.64$ & $0.0281$ & {\color{red!90!black}$\mathbf{0.0000}$} & $0.5247$ & $0.8930$ & $2.6603$ & $1.98$ & $5.41$ &\textcolor[HTML]{FF7E79}{\xmark}\\ 
\bottomrule
\end{tabular}
}
\label{tab:paramscaling_left}
\end{table}

\section{Experimental Details of \ttsmethod}
\label{app:tts}

\paragraph{Experimental Setup}

The formal description of \ttsmethod is provided in \Cref{alg:tts}. Building upon this framework, we evaluate \ttsmethod on six diverse in silico protein design tasks, encompassing structure-guided, biophysical, and functional objectives. Unless otherwise noted, all experiments share a unified optimization configuration: 10 total iterations ($T = 10$), 50 candidates(batch size) generated per round ($N = 50$), and top-$k$ selection size $k = 5$. 

\paragraph{Task-Specific Settings}

Each task is implemented with a distinct evaluation metric and verifier model. \Cref{tab:task-config} summarizes the datasets used, verifier models, and key hyperparameters. For each task, we report mean top-$k$ from cumulative generated proteins.

\begin{table}[h]
\centering
\caption{Summary of task configurations and hyperparameters.}
\label{tab:task-config}
\resizebox{\textwidth}{!}{  
\begin{tabular}{lllll}
\toprule
\textbf{Task} & \textbf{Input Samples} & \textbf{Verifier Model} & \textbf{Inference Steps} & \textbf{Initial Noise Level} \\
\midrule
Orphan Design & T1026, T1033, T1039, T1043, T1064, T1074 & TM-score (ESMFold) & 50 & 0.99 \\
General Design & 8JGO\_A, 8SW5\_C, 8WOP\_A, 8XJE\_B, 9B1R\_A & TM-score (ESMFold) & 50 & 0.90 \\
Optimal Temperature & an in-house protein & Seq2Topt~\cite{qiu2025seq2topt} & 250 & 0.99 \\
Optimal pH & an in-house protein & Seq2pHopt~\cite{qiu_seq2topt_2023} & 250 & 0.99 \\
Enzyme EC Matching & F2XF95, F2XF96, F2XFA8, R9QMQ9, R9QMR0, R9QMR3, R9QMW1, R9QMW5, R9QMY8, R9QMY9 & CLEAN~\cite{yu2023enzyme} & 50 & 0.99 \\
Reaction Activity & A0A0A7E972, B7X755, O42275, P18142, P32749 & CLIPZyme~\cite{mikhael2024clipzyme} & 100 & 0.99 \\
\bottomrule
\end{tabular}
}
\end{table}
\paragraph{Reward Function Definitions}

The reward function $R(\cdot)$ is defined separately for each task to reflect task-specific design objectives. For the Orphan Design and General Design tasks, $R$ is the TM-score between the ESMFold-predicted structure of a generated sequence and the known structure of the input wild-type protein. For the Optimal Temperature and Optimal pH tasks, the design objective is to match target physiological conditions—specifically 37°C and pH 7, respectively. In these cases, $R(x) = -|f(x) - y^*|$, where $f(x)$ is the output of the predictor (Seq2Topt or Seq2pHopt) and $y^*$ is the target value. For the Enzyme EC Matching task, the initial inputs are proteins predicted by CLEAN to belong to EC class 4.2.3.120, and the goal is to evolve sequences toward EC class 4.2.3.113. Here, $R(x)$ is the classification confidence assigned by CLEAN to the target EC class. Finally, for the Reaction Activity task, the goal is to maximize the predicted catalytic competence toward the target reaction \ce{[CH3][N+](CH3)(CH3)CH2CHO + H2O + O2 -> [CH3][N+](CH3)(CH3)CH2COOH + H2O2} as evaluated by CLIPZyme, and $R(x)$ is defined as the model's confidence score for this specific transformation.

\paragraph{Baseline Adaptation}

To ensure a fair comparison with baseline methods (ALDE, EVOLVEpro, MLDE), we modify their original frameworks by simply replacing their internal fitness predictors or experimental feedback modules with a shared external verifier oracle $R(\cdot)$. This harmonization ensures that all methods receive identical feedback signals across evolution process, isolating the effect of optimization strategies from feedback quality.

For each baseline, we simulate directed evolution over total 500 variants. After each round(50 proteins are considered as a round), we compute the cumulative top-$k$ mean score using the shared verifier, mirroring the evaluation protocol used for \ttsmethod. This allows us to compare the evolution trajectories and final design quality under a consistent verification budget and objective.

\begin{algorithm}[t]
\caption{\ttsmethod}
\label{alg:tts}
\begin{algorithmic}[1]
\Require Initial MSA $\mX$, batch size $N$, number of rounds $T$, selection size $k$, external verifier $R(\cdot)$
\Ensure Top-$k$ sequences from full generation history, ranked by verifier

\State $\mP^{(0)} \gets \text{BuildProfile}(\mX)$ \Comment{Convert MSA to profile}
\State $\mathcal{H} \gets \emptyset$ \Comment{Initialize history of generated sequences}

\For{$t = 0$ \textbf{to} $T-1$}
\State $\{\mathbf{x}_i^{(t)}\}_{i=1}^N \gets \text{\method}(\mP^{(t)})$ \Comment{Generate $N$ sequences conditioned on profile}

\For{$i = 1$ \textbf{to} $N$}
\State $s_i^{(t)} \gets R(\mathbf{x}_i^{(t)})$ \Comment{Compute score}
\State $\mathcal{H} \gets \mathcal{H} \cup {(\mathbf{x}_i^{(t)}, s_i^{(t)})}$ \Comment{Add to generation history}
\EndFor
\State $\mathcal{X}^{(t)} \gets \text{Top-}k(\mathcal{H})$ \Comment{Select top-$k$ sequences so far}
\State $\mP^{(t+1)} \gets \text{BuildProfile}(\mathcal{X}^{(t)})$ \Comment{Update profile from selected sequences}
\EndFor

\State \Return $\text{Top-}k(\mathcal{H})$ \Comment{Final output: best sequences across all rounds}
\end{algorithmic}
\end{algorithm}

\section{\ttsmethod Ablation Study}
\label{tts_ablation}
\subsection{Assessing the Role of the Generative Model in Test-Time Scaling}
To assess the necessity of using a pretrained protein generative model in \ttsmethod, we conduct an ablation study in which we replace \method with a simple stochastic generator based on random mutation. Specifically, at each round of test-time scaling, we randomly perturb 10\% of the residues in the top-$1$ sequences, uniformly sampling new amino acids at mutated positions. This generates a new candidate pool without relying on a trained protein foundation model. All other components of the TTS loop—including the verifier, profile update mechanism, and sampling procedure—remain unchanged.

\begin{figure}[htb]
    \centering
    \includegraphics[width=1\linewidth]{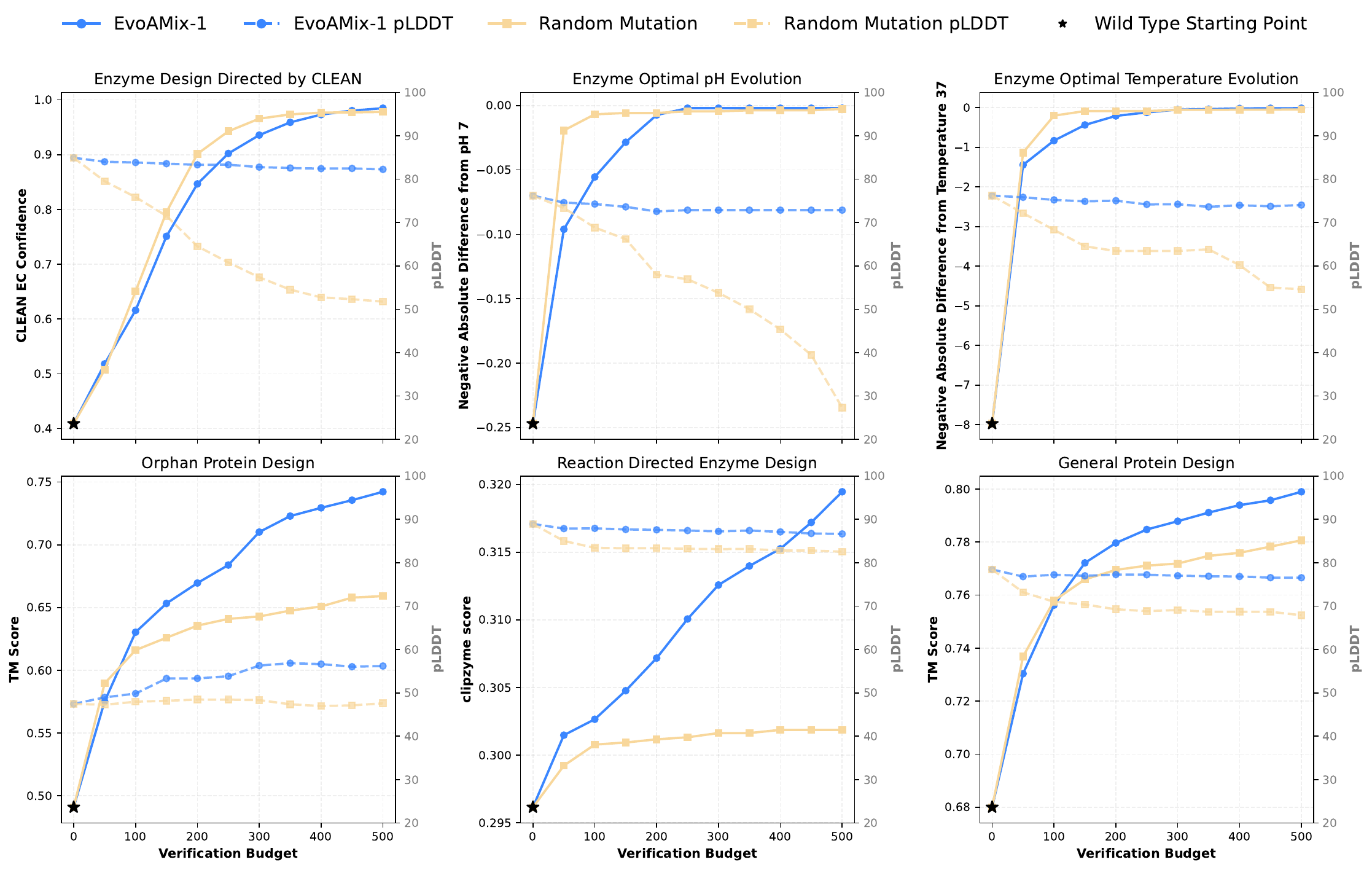}
    \caption{Ablation study: structural integrity of sequences generated by \ttsmethod versus random mutation.
    Each curve shows the trajectory of the task reward (solid line, left y-axis) and mean pLDDT (dashed line, right y-axis) over increasing verifier calls. While random mutation achieves comparable or even higher reward scores in some tasks, it causes a rapid decline in structural confidence. In contrast, \ttsmethod maintains consistently high pLDDT values throughout optimization, despite the absence of any explicit structure-based supervision.
    }
    \label{fig:ablation_pLDDT}
\end{figure}

\Cref{fig:ablation_pLDDT} compares the performance of this random mutation variant against the full \ttsmethod across six directed evolution benchmarks. On certain objectives—such as optimal temperature and pH—random mutation initially shows comparable or even slightly better performance than \ttsmethod in terms of the target reward metric. However, this superficial gain comes at a critical cost: the predicted structural quality (measured by pLDDT) of generated proteins degrades sharply over time. By contrast, \ttsmethod maintains consistently high pLDDT scores throughout the optimization process, despite not receiving any explicit structural feedback during evolution.

\begin{figure}[htb]
    \centering
    \includegraphics[width=1\linewidth]{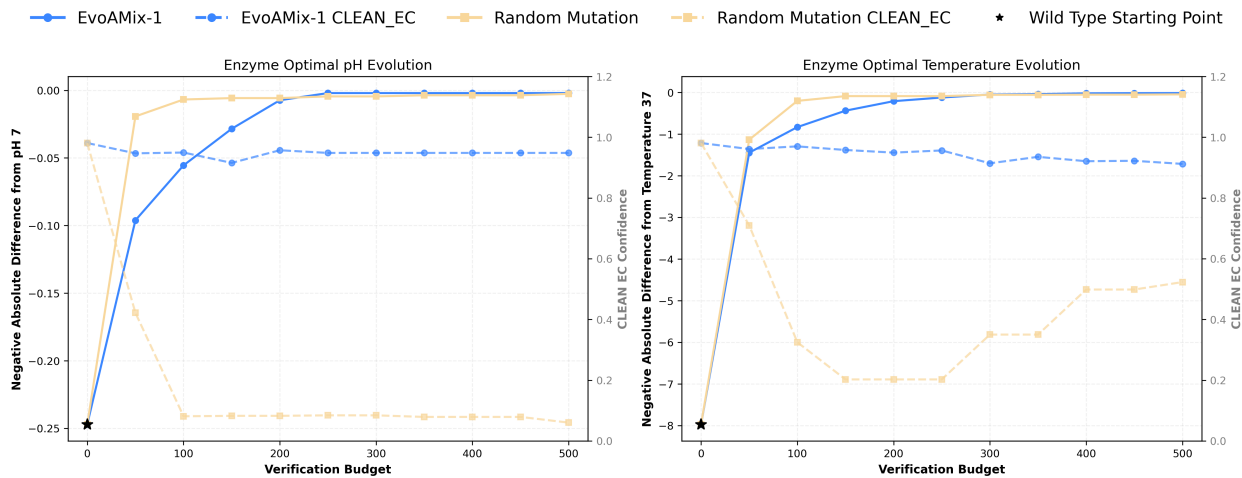}
    \caption{
    Ablation study: functional consistency of generated sequences under pH and temperature optimization.
    We measure CLEAN-predicted EC confidence scores (y-axis) over increasing verifier calls. \ttsmethod consistently retains the enzymatic identity of the starting protein (i.e., original EC number), even without using CLEAN as a part of reward signal. In contrast, random mutation degrades functional consistency over time, indicating that it drifts away from the original functional class.
    }
    \label{fig:ablation_ec}
\end{figure}

To further probe the functional integrity of generated proteins, we evaluated whether the optimized sequences retained the original enzymatic function class, as measured by CLEAN confidence scores with respect to the input EC number. \Cref{fig:ablation_ec} shows the CLEAN prediction scores during pH and temperature optimization. While \ttsmethod preserves near-perfect EC predictions across all iterations, random mutation causes CLEAN confidence to deteriorate progressively—indicating a loss of functional specificity. This difference is particularly notable given that no CLEAN-based supervision was used in either setting, further underscoring the inductive bias encoded in the \method generative prior.

These results highlight that naïve mutation-based exploration—even when embedded in a TTS framework—can rapidly drift away from biologically plausible regions of sequence space. In contrast, the structured sampling behavior of \method ensures that sequence modifications remain compatible with both structure and function constraints. We conclude that a pretrained foundation model is essential for ensuring biological fidelity during test-time scaling.

\subsection{Assessing the Impact of Iterative Condition Refinement}
\begin{figure}[ht]
    \centering
    \includegraphics[width=1\linewidth]{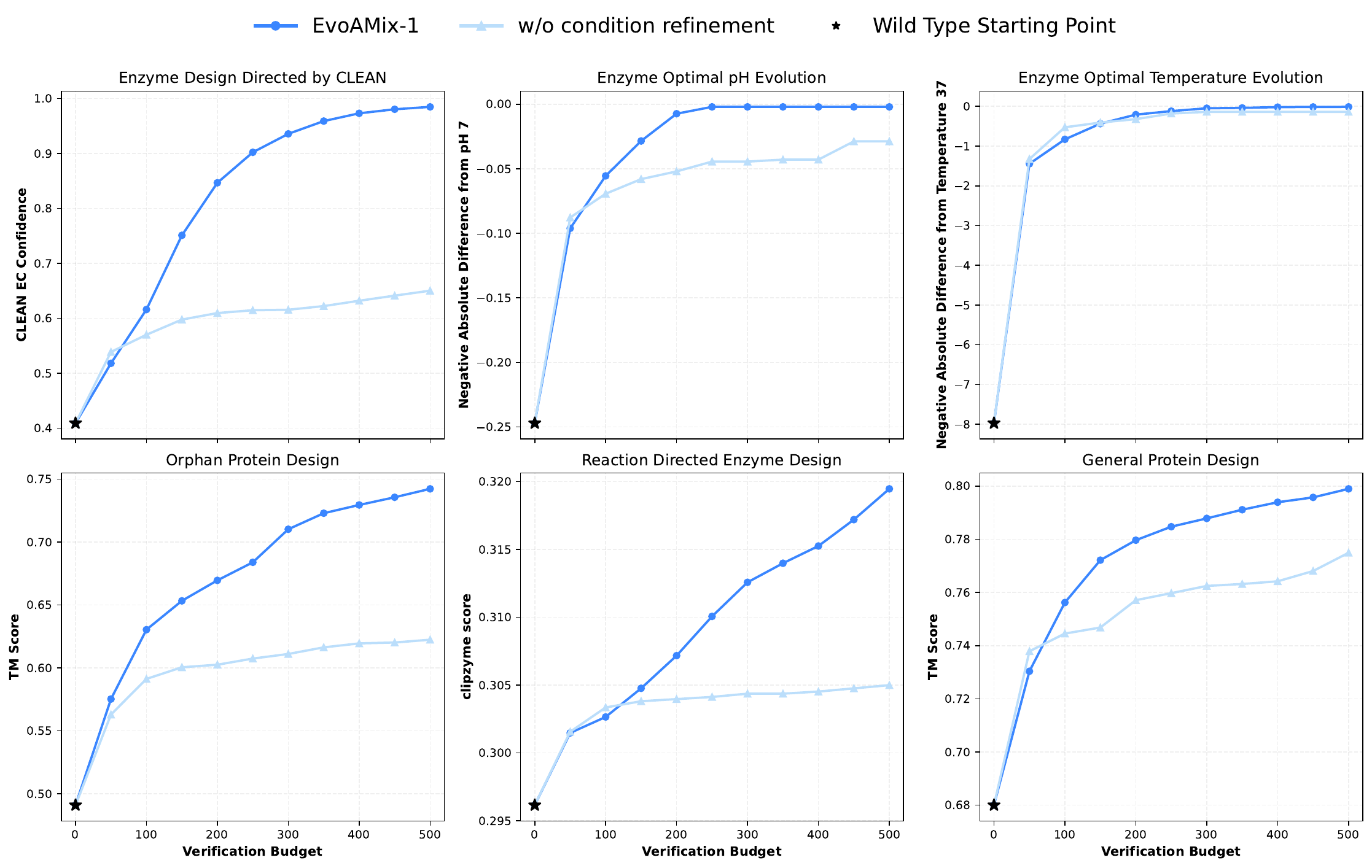}
    \caption{
    Ablation study: removing iterative profile refinement during test-time scaling.
    In this variant, \ttsmethod is executed without updating the profile $\mP^{(t)}$ across rounds—i.e., all generations are conditioned on the initial profile $\mP^{(0)}$ derived from the wild-type sequence. Across all tasks, this non-adaptive variant leads to significantly worse final performance and slower convergence, highlighting the essential role of iterative prompt refinement in guiding the generation process toward regions of higher fitness.
    }
    \label{fig:ablation_refinement}
\end{figure}
To isolate the impact of iterative condition adaptation in \ttsmethod, we conduct an ablation in which the profile prompt $\mP^{(t)}$ is fixed throughout the entire optimization process. Specifically, after computing the initial profile $\mP^{(0)}$ from the wild-type MSA, all subsequent generations are conditioned on this static prompt, regardless of the quality of sequences discovered during evolution. In other words, we remove the feedback loop between the verifier and the generative model, breaking the propose–verify–refine cycle.

\Cref{fig:ablation_refinement} summarizes the result across all six design tasks. Compared to full \ttsmethod, the fixed-profile variant consistently underperforms, both in terms of final reward and convergence speed. These results demonstrate that iterative prompt refinement is a key driver of performance gains in TTS: by updating the proposal distribution to reflect high-quality sequences discovered so far, the model is progressively steered toward regions of sequence space favored by the verifier. Without this adaptive mechanism, the optimization stagnates, highlighting the importance of dynamic conditioning even in the absence of parameter updates.

\section{Verification in wet-lab}
\label{app:wet-lab}

As silico experiments has shown \method's effectiveness in protein-direct evolution and test-time scaling. In this section, wet-lab experiments are included to validate such a generative model to produce functional proteins. We selected the AmeR protein as a target and aimed to optimize its DNA-binding affinity. We employed the model to generate a library of AmeR mutants, whose fitness was subsequently verified through wet-lab experiments. 

The results demonstrated that model-generated mutants enhanced the protein's activity by up to 50-fold compared to the wild type. This represents an approximate 77\% improvement over the best previously reported results. These findings confirm the model's ability to reliably generate novel, functional protein sequences. This capability allows for the simulation of directed evolution, which can significantly accelerate the experimental process while reducing the associated costs and labor.

\paragraph{Wet-lab Backgrounds.}

    The AmeR protein, a member of the TetR family of transcriptional regulators, is central to this study. In synthetic biology, AmeR is widely utilized as an orthogonal biological logic gate, where it represses gene expression by binding to its corresponding promoter\citep{stanton2014genomic}. A key challenge in engineering AmeR is the scarcity of its homologs, which results in a vast and underexplored mutational landscape, complicating targeted functional optimization.

    This study aims to employ our generative model to produce AmeR variants with enhanced DNA-binding affinity. These improved variants are envisioned as more reliable molecular switches for constructing complex and predictable synthetic biology systems, with potential applications in metabolic engineering, biosensors, and cell therapy.


\paragraph{Wet-lab Methods.}
\label{exp:fold_propa}


    To comprehensively explore the sequence-function landscape of AmeR and identify mutants with enhanced fitness, we employed our generative model, \method. The model was conditioned on a profile constructed from the wild-type AmeR sequence, its multiple sequence alignment (MSA), and a set of previously characterized variants. Using the 650M-parameter version of our model (\method-650M), we performed conditional sampling. The generated sequences were then subjected to similarity filtering based on specific hyper-parameter settings, yielding a final set of 40 candidate sequences, each containing ten or fewer mutations relative to the wild-type.

    To quantify the DNA-binding affinity of AmeR mutants, we established a fluorescent reporter assay adapted from EvoAI\citep{ma2025evoai} in the wet lab. In this system, the DNA-binding activity of AmeR is inversely coupled to the expression of an enhanced yellow fluorescent protein (eYFP). Specifically, high-affinity binding of an AmeR variant to its corresponding operator (pAmeR) represses eYFP transcription, leading to a low fluorescence signal. Conversely, weak or no binding results in high eYFP expression and a strong signal. The expression of the AmeR variants was placed under the control of an IPTG-inducible pTac promoter. We quantified the repressive strength using the metric ``fold repression'' calculated as follows:

    \begin{equation}
        \mathbf{Fold \ Repression} = \frac{\text{Fluorescence}_{[\mathbf{uninduced}]} - \text{Fluorescence}_{[\mathbf{background}]} } {\text{Fluorescence}_{[\mathbf{induced}]} - \text{Fluorescence}_{[\mathbf{background}]} }
    \end{equation}

    Here, background fluorescence was measured from a control strain carrying an empty plasmid. This metric is positively correlated with the DNA-binding affinity of AmeR, representing its efficacy as a transcriptional ``off-switch''.

    To acquire the data, engineered strains harboring each mutant were cultured, and the fluorescence signal was measured via flow cytometry to determine the fold repression for each variant and the wild type.

\paragraph{Baselines.}The performance of our model-generated mutants was evaluated against three key baselines:

\begin{itemize}
    \item \textbf{Wild-Type (WT)}: The original, unmodified native AmeR sequence.
    \item \textbf{Single-Mutant Library}: A collection of previously characterized AmeR variants, each containing a single amino-acid substitution relative to the wild-type.
    \item \textbf{EvoAI Variants}: Mutants generated by EvoAI\citep{ma2025evoai}, a previously reported evolutionary method. This method identifies beneficial single mutations (anchors) from local fitness peaks and then combinatorially introduces them to create new sequences with seven or fewer substitutions.
\end{itemize}

\paragraph{Wet-lab results.}
Our results indicate that \method explores the sequence space more effectively than existing methods. As illustrated by our experimental pipeline in \Cref{fig:wetlab_path}, \method enables the controlled introduction of both single and multiple mutations. A summary of the peak fold repression values achieved by all methods is presented in \Cref{fig:wetlab_result}. Notably, \method-designed variants demonstrated a significant improvement over all baselines. The top-performing \method variant, which contained 10 mutations, exhibited a fold repression nearly double that of the best EvoAI variant (7 mutations). 
    
These findings underscore \method's superior ability to navigate the sequence-function landscape to identify high-impact mutations and discover highly active variants.

\end{document}